# Wave motions in unbounded poroelastic solids infused with compressible fluids

S. Quiligotti, G.A. Maugin and F. dell'Isola

*This paper is dedicated to the memory of Professor Eugen Soós*

**Abstract.** Looking at rational solid-fluid mixture theories in the context of their biomechanical perspectives, this work aims at proposing a two-scale constitutive theory of a poroelastic solid infused with an inviscid compressible fluid. The propagation of steady-state harmonic plane waves in unbounded media is investigated in both cases of unconstrained solid-fluid mixtures and fluid-saturated poroelastic solids. Relevant effects on the resulting characteristic speed of longitudinal and transverse elastic waves, due to the constitutive parameters introduced, are finally highlighted and discussed.

## 1. Introduction

A solid-fluid mixture is customarily thought of as a couple of body manifolds embedded into the three-dimensional Euclidean space (see e.g. Atkin and Craine [2], Bowen [6], Krishnaswamy and Batra [21], Rajagopal and Tao [28], Truesdell [34, 35]), so as to occupy a common smooth region of the physical environment while undergoing independent motions, here assumed to take place in a neighborhood of the initial solid configuration (section 2).

If neither chemical reactions nor phase transitions occur, the mass of each constituent is conserved along the corresponding motion. Moreover, the local form of the fluid-mass conservation law can be advantageously written with respect to the initial configuration, also regarded as a reference configuration (section 3).

In order to take into consideration, at least coarsely, the most remarkable microstructural properties of the mixture (see e.g. Schrefler [30]), the concept of volume fraction (cf. Bowen [5, 7], de Boer [9, 10], Wilmanski [38]) is furthermore introduced within the framework of a first-order gradient theory (sections 5.1–5.2), assuming a linear constitutive dependence of *microscopic* mass densities on *macroscopic* kinematical descriptors (section 4).

If the saturation constraint is fulfilled (Klisch [20], Svendsen and Hutter [31]), i.e. the volume occupied by the constituents equals the volume available to the mixture, then the stress response is determined by the motion except for an arbitrary contribution, due to the saturation pressure which arises in the material so as to maintain each constituent in contact with the other one. Such a pressure can be truly regarded as a Lagrangian multiplier in the expression of the strain-energy density per unit volume of the mixture (dell'Isola et al. [11], Quiligotti et al. [13]), so as to deduce the splitting rule which governs the distribution of such a pressure among the constituents as a result of the theory (section 5.2).

As far as the power expended by inertial forces is concerned, the overall kinetic energy density per unit reference volume of the mixture is defined as the sum of peculiar kinetic energy densities. Its material derivative, following the motion of the mixture as a whole, is required to equal the power expended by inertial forces (section 5.3). This assumption, physically motivated although far from evident, seems to corroborate the importance of coupled inertial interactions (Biot [3], cf. de Boer [10]).

The principle of virtual power (see e.g. Germain [16, 17]) is finally used as a main tool (Di Carlo [12], Maugin [23]) to deduce the set of balance equations and boundary conditions that governs the nonlinear dynamics of the mixture (Quiligotti et al. [13]). With the aim of investigating the propagation of elastic plane waves in unbounded media (Achenbach [1], Graff [18]), these equations have been linearized (section 5.4), and their harmonic steady-state solutions have been found in both cases of unconstrained solid-fluid mixtures and fluid-satured porous solids (section 5.5). A few relevant remarks, inherently based on a comparative analysis of emerging results, complete this study and implicitly outline some possible biomechanical applications that may be envisaged as a further development of the proposed two-scale constitutive theory (cf. Cowin [8]; see also Humphrey and Rajagopal [19], Ehlers and Markert [15], Tao et al. [33], Taber [32]).

## 2. Kinematics

Let us consider a binary solid-fluid mixture, consisting of two smooth three-dimensional material manifolds, denoted by $\mathfrak{B}_{\mathrm{S}}$ and $\mathfrak{B}_{\mathrm{F}}$ (figure 1).

By assumption (Noll and Virga [26]), a time-independent smooth embedding of the solid body into the physical Euclidean space $\mathcal{E}$,

$$\mathcal{K}_{\mathrm{S}} : \mathfrak{B}_{\mathrm{S}} \to \mathcal{E} \tag{1}$$

$$\mathfrak{X}_{\mathrm{S}} \mapsto \mathbf{X}, \tag{2}$$

associates any material solid particle with a reference place.

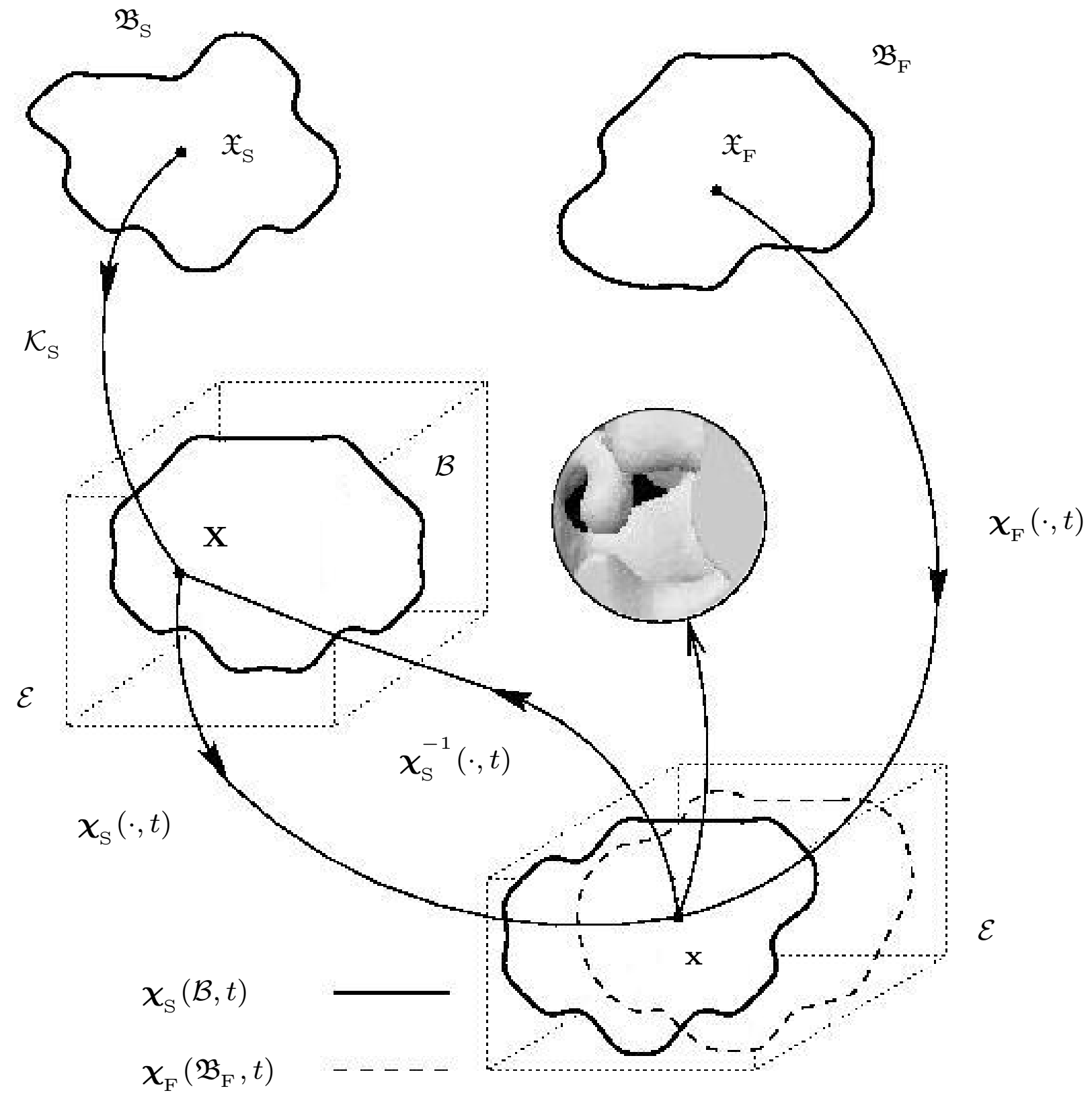

Figure 1. Kinematics of a binary solid-fluid mixture.

A smooth S-motion† may be described as a time sequence of mappings,

$$\boldsymbol{\chi}_S(\cdot, t): \ \mathcal{B} \rightarrow \mathcal{E} \tag{3}$$

$$\mathbf{X} \mapsto \mathbf{x}, \tag{4}$$

which carry any solid particle from its reference place to its current one.

In particular, so as to linearize the equations that govern the dynamics of the mixture, we shall focus our attention on a smooth 1-parameter family of S-motions,

$$\boldsymbol{\chi}_S(\mathbf{X}, t) := \mathbf{X} + \varepsilon \boldsymbol{u}_S(\mathbf{X}, t), \tag{5}$$

† For the sake of conciseness, we shall refer to any motion of the body manifold $\mathfrak{B}_\alpha$ (with $\alpha \in \{S, F\}$ ) as an *$\alpha$-motion.*

whose gradient straightforwardly results in:

$$\operatorname{Grad}\boldsymbol{\chi}_{\mathrm{S}} = \mathbf{I} + \varepsilon \operatorname{Grad}\boldsymbol{u}_{\mathrm{S}} \,. \tag{6}$$

Similarly, a smooth F-motion can be described as a time sequence of embeddings which map the fluid-body manifold onto its current shape,

$$\boldsymbol{\chi}_{\mathrm{F}}(\cdot,t) : \ \mathfrak{B}_{\mathrm{F}} \ \rightarrow \ \mathcal{E} \tag{7}$$

$$\mathfrak{X}_{\mathrm{F}} \ \mapsto \ \mathbf{x}\,, \tag{8}$$

so that, at time $t$, any given place in the current shape of the mixture,

$$x = \boldsymbol{\chi}_{\mathrm{S}}(\mathcal{K}_{\mathrm{S}}(\mathfrak{X}_{\mathrm{S}}),t) = \boldsymbol{\chi}_{\mathrm{F}}(\mathfrak{X}_{\mathrm{F}},t) \in \mathcal{B}(t) \,, \tag{9}$$

with

$$\mathcal{B}(t) := \Big\{\boldsymbol{\chi}_{\mathrm{S}}(\mathcal{B},t) \bigcap \boldsymbol{\chi}_{\mathrm{F}}(\mathfrak{B}_{\mathrm{F}},t)\Big\} \quad \text{and} \quad \mathcal{B} = \mathcal{K}_{\mathrm{S}}(\mathfrak{B}_{\mathrm{S}}) \,, \tag{10}$$

is simultaneously occupied by a pair of different material particles, $\mathfrak{X}_{\mathrm{S}} \in \mathfrak{B}_{\mathrm{S}}$ and $\mathfrak{X}_{\mathrm{F}} \in \mathfrak{B}_{\mathrm{F}}$.

As the reference shape $\mathcal{B}$ does not depend on time, the velocity of any $\alpha$-particle, $\mathfrak{X}_\alpha \in \mathfrak{B}_\alpha$, is just given by the material derivative following the $\alpha$-motion,

$$\mathbf{v}_{\mathrm{S}}(\mathbf{x},t) := \left.\frac{\partial \boldsymbol{\chi}_{\mathrm{S}}(\mathbf{X},\tau)}{\partial \tau}\right|_{\tau=t} \tag{11}$$

$$\mathbf{v}_{\mathrm{F}}(\mathbf{x},t) := \left.\frac{\partial \boldsymbol{\chi}_{\mathrm{F}}(\mathfrak{X}_{\mathrm{F}},\tau)}{\partial \tau}\right|_{\tau=t}, \tag{12}$$

where $\mathbf{X} = \mathcal{K}_{\mathrm{S}}(\mathfrak{X}_{\mathrm{S}})$ and $\mathbf{x} = \boldsymbol{\chi}_{\mathrm{S}}(\mathbf{X},t) = \boldsymbol{\chi}_{\mathrm{F}}(\mathfrak{X}_{\mathrm{F}},t)$.

In particular, the kinematical assumption (5) leads the Eulerian velocity field of the solid body (11) to read

$$\mathbf{v}_{\mathrm{S}}(\mathbf{x},t) = \varepsilon \left.\frac{\partial\,\boldsymbol{u}_{\mathrm{S}}(\mathbf{X},\tau)}{\partial \tau}\right|_{\tau=t} =: \varepsilon \boldsymbol{v}_{\mathrm{S}}(\mathbf{X},t)\,. \tag{13}$$

Moreover, as we exclude a priori the possibility that any three-dimensional region of the reference shape collapses under the motion $\boldsymbol{\chi}_{\mathrm{S}}$,

$$\mathrm{J}_{\mathrm{S}}(\mathbf{x},t) = \det \mathbf{F}_{\mathrm{S}}(\mathbf{X},t) > 0\,, \quad \text{with} \quad \mathbf{F}_{\mathrm{S}} := \operatorname{Grad}\boldsymbol{\chi}_{\mathrm{S}}\,, \tag{14}$$

there exists a smooth inverse mapping,

$$\boldsymbol{\chi}_{\mathrm{S}}^{-1}(\cdot,t) : \boldsymbol{\chi}_{\mathrm{S}}(\mathcal{B},t) \rightarrow \mathcal{E}\,, \tag{15}$$

that satisfies the trivial identity

$$\mathbf{X} = \boldsymbol{\chi}_{\mathrm{S}}^{-1}\big(\boldsymbol{\chi}_{\mathrm{S}}(\mathbf{X},t),t\big)\,, \quad \forall\,\mathbf{X} \in \mathcal{B}\,, \tag{16}$$

yielding the following property at any time $t$ and place $\mathbf{X}$,

$$\operatorname{grad}\boldsymbol{\chi}_{\mathrm{S}}^{-1}(\mathbf{x},t) = \big(\operatorname{Grad}\boldsymbol{\chi}_{\mathrm{S}}(\mathbf{X},t)\big)^{-1}. \tag{17}$$

In order to describe the motion of the fluid constituent through the reference shape of the solid, we also notice that any fluid particle interacts with a smooth 1-parameter family of solid ones, moving along the curve

$$\boldsymbol{\chi}_{\mathrm{S}}^{-1}\left(\boldsymbol{\chi}_{\mathrm{F}}(\mathfrak{X}_{\mathrm{F}},\cdot),\cdot\right):\ t \mapsto \mathbf{X}, \tag{18}$$

at the velocity $\mathbf{w}_{\mathrm{F}}(\mathbf{X},t)$, defined by the relation

$$\mathbf{v}_{\mathrm{F}}(\mathbf{x},t) = \mathbf{F}_{\mathrm{S}}(\mathbf{X},t)\,\mathbf{w}_{\mathrm{F}}(\mathbf{X},t) + \mathbf{v}_{\mathrm{S}}(\mathbf{x},t), \tag{19}$$

where, as usual, $\mathbf{x} = \boldsymbol{\chi}_{\mathrm{F}}(\mathfrak{X}_{\mathrm{F}},t) = \boldsymbol{\chi}_{\mathrm{S}}(\mathbf{X},t)$.

With the aim of linearizing the equations that govern the nonlinear dynamics of the mixture, we shall consider a smooth 1-parameter family of relative velocity fields,

$$\mathbf{w}_{\mathrm{F}} := \boldsymbol{w}_{\mathrm{FO}} + \gamma\,\boldsymbol{w}_{\mathrm{F}}, \tag{20}$$

assuming, for the sake of simplicity, that the two independent perturbation parameters introduced so far are of the same order of magnitude, $o(\gamma) = o(\varepsilon)$, and the unperturbed velocity field $\boldsymbol{w}_{\mathrm{FO}}$ is identically equal to the null vector,

$$\boldsymbol{w}_{\mathrm{FO}}(\mathbf{X},t) = \mathbf{0}, \quad \forall\,\mathbf{X}\in\mathcal{B}, \tag{21}$$

so as to get the following expression for the Eulerian fluid velocity (19):

$$\mathbf{v}_{\mathrm{F}} \overset{\circ}{=} \varepsilon\boldsymbol{v}_{\mathrm{S}} + \varepsilon\boldsymbol{w}_{\mathrm{F}} =: \varepsilon\boldsymbol{v}_{\mathrm{F}}. \tag{22}$$

## 3. Mass conservation

By definition, the $\alpha$-mass content of any smooth region of the current shape of the solid-fluid mixture, $\mathcal{V}\subset\mathcal{B}(t)$, is given by the measure

$$\mathcal{M}_{\alpha}(\mathcal{V}) = \int_{\mathcal{V}} \varrho_{\alpha}, \quad \text{with } \alpha\in\{\mathrm{S},\mathrm{F}\}. \tag{23}$$

If mixture constituents undergo neither chemical reactions nor phase transitions, the time derivative of the $\alpha$-mass content of $\mathcal{V}$ has to vanish following the $\alpha$-motion of the migrating material surface that coincides, at the given time $\tau = t$, with the boundary of the chosen smooth region $\mathcal{V}$, i.e.

$$\left\{\frac{d}{d\tau}\int_{\mathcal{V}_{\alpha}(\tau)} \varrho_{\alpha}\right\}_{\tau=t} = \int_{\mathcal{V}}\left(\frac{\partial\varrho_{\alpha}}{\partial t} + \operatorname{div}(\varrho_{\alpha}\mathbf{v}_{\alpha})\right) = 0, \tag{24}$$

leading to the local mass conservation law

$$\frac{\partial\varrho_{\alpha}}{\partial t} + \operatorname{div}(\varrho_{\alpha}\mathbf{v}_{\alpha}) = 0. \tag{25}$$

Introducing an $\alpha$-mass density per unit reference volume of the mixture,

$$\varrho_{\alpha}^{\star}(\mathbf{X},t) = \mathrm{J}_{\mathrm{S}}(\mathbf{x},t)\,\varrho_{\alpha}(\mathbf{x},t), \quad \forall\,\mathbf{x} = \boldsymbol{\chi}_{\mathrm{S}}(\mathbf{X},t)\in\mathcal{B}(t), \tag{26}$$

as the scalar field on $\mathcal{B}$ that satisfies the integral equality

$$\int_{\mathcal{V}^\star} \varrho_\alpha^\star := \int_{\mathcal{V}} \varrho_\alpha \,, \qquad \forall\, \mathcal{V}^\star = \boldsymbol{\chi}_{\mathrm{S}}^{-1}(\mathcal{V}, t)\,, \tag{27}$$

we may also evaluate, at any given time $\tau = t$, the material derivative of the $\alpha$-mass content of $\mathcal{V}$, following the motion of the $\alpha$-th constituent through the reference shape $\mathcal{B}$, and taking into account that $\mathcal{V}_\alpha^\star(\tau) = \boldsymbol{\chi}_{\mathrm{S}}^{-1}(\mathcal{V}_\alpha(\tau), \tau)$, with $\mathcal{V}_\alpha(t) = \mathcal{V}$,

$$\left\{ \frac{d}{d\tau} \int_{\mathcal{V}_\alpha^\star(\tau)} \varrho_\alpha^\star \right\}_{\tau = t} = \int_{\mathcal{V}_\alpha^\star(t)} \left( \frac{\partial \varrho_\alpha^\star}{\partial t} + \mathrm{Div}\,(\varrho_\alpha^\star \mathbf{w}_\alpha) \right) = 0\,, \tag{28}$$

so as to get an alternative form of the local mass conservation law,

$$\frac{\partial \varrho_\alpha^\star}{\partial t} + \mathrm{Div}\,(\varrho_\alpha^\star \mathbf{w}_\alpha) = 0\,. \tag{29}$$

Expanding in power series the reciprocal of the smooth function

$$\mathrm{J}_{\mathrm{S}}(\mathbf{x}, t) = \det \mathbf{F}_{\mathrm{S}}(\mathbf{X}, t)\,, \quad \forall\, \mathbf{x} = \boldsymbol{\chi}_{\mathrm{S}}(\mathbf{X}, t) \in \mathcal{B}(t) \tag{30}$$

in a neighborhood of the reference shape,

$$\frac{1}{\det \mathbf{F}_{\mathrm{S}}} = 1 - \varepsilon\, \mathrm{Div}\, \boldsymbol{u}_{\mathrm{S}} + o(\varepsilon)\,, \tag{31}$$

we deduce that, by virtue of definition (26), the value of the current mass density of any $\alpha$-particle that occupies the given place $\mathbf{x}$ at time $t$,

$$\varrho_\alpha(\mathbf{x}, t) = \varrho_\alpha^\star(\mathbf{X}, t)\,(1 - \varepsilon\, \mathrm{Div}\, \boldsymbol{u}_{\mathrm{S}}(\mathbf{X}, t)) + o(\varepsilon)\,, \quad \text{with} \;\; \mathbf{x} = \boldsymbol{\chi}_{\mathrm{S}}(\mathbf{X}, t)\,, \tag{32}$$

equals the value of the reference $\alpha$-mass density $\varrho_\alpha^\star(\mathbf{X}, t)$ if the displacement field of the solid body results to be divergence-free at the same given time and place, i.e. $\mathrm{Div}\; \boldsymbol{u}_{\mathrm{S}}(\mathbf{X}, t) = 0$.

As the reference shape does not depend on time, when $\alpha = \mathrm{S}$ the integral conservation laws (24) and (28) lead to the trivial conclusion that the scalar field $\varrho_{\mathrm{S}}^\star$ is independent of time ( $\mathbf{w}_{\mathrm{S}} = \mathbf{0}$ ) as well,

$$\frac{\partial \varrho_{\mathrm{S}}^\star}{\partial t} = 0\,, \tag{33}$$

and thus, bearing in mind the equality (32),

$$\varrho_{\mathrm{S}}(\mathbf{x}, t) - \varrho_{\mathrm{S}}^\star(\mathbf{X}) \overset{\circ}{=} -\varrho_{\mathrm{S}}^\star(\mathbf{X})\; \mathrm{Div}\, \varepsilon \boldsymbol{u}_{\mathrm{S}}(\mathbf{X}, t)\,, \quad \forall\, \mathbf{x} = \boldsymbol{\chi}_{\mathrm{S}}(\mathbf{X}, t) \in \mathcal{B}(t)\,. \tag{34}$$

Because of the overlapping between the two constituents, any smooth region of the current shape of the mixture can also be associated with a fluid subbody. In particular, for any choice of $\mathcal{V}_{\mathrm{S}}(t)$, there exists a fluid subbody, $\mathfrak{P}_{\mathrm{F}} \subset \mathfrak{B}_{\mathrm{F}}$, such that $\mathcal{V}_{\mathrm{F}}(\tau) = \boldsymbol{\chi}_{\mathrm{F}}(\mathfrak{P}_{\mathrm{F}}, \tau)$, with $\mathcal{V}_{\mathrm{F}}(t) = \mathcal{V}_{\mathrm{S}}(t)$ at time $\tau = t$.

In order to linearize the local expression of the $\alpha$-mass balance (29) in the case of $\alpha = \mathrm{F}$, we shall assume that the fluid motion takes place in a neighborhood of the reference shape of the mixture, i.e.

$$\varrho^\star_{\mathrm{F}}(\mathbf{X},t) \;=\; \varrho^\star_{\mathrm{F0}}(\mathbf{X}) + \varepsilon\varrho^\star_{\mathrm{F1}}(\mathbf{X},t) + o\,(\varepsilon)\,, \tag{35}$$

$$\mathbf{w}_{\mathrm{F}}(\mathbf{X},t) \;=\varepsilon \boldsymbol{w}_{\mathrm{F}}(\mathbf{X},t)\;. \tag{36}$$

As a consequence, it is possible to deduce that the F-mass density fields $\varrho^\star_{\mathrm{F0}}$ and $\varrho^\star_{\mathrm{F1}}$ have to fulfill, respectively, two independent requirements:

$$\frac{\partial \varrho^\star_{\mathrm{F0}}}{\partial t} = 0\,, \tag{37}$$

$$\frac{\partial \varrho^\star_{\mathrm{F1}}}{\partial t} + \mathrm{Div}\left(\varrho^\star_{\mathrm{F0}}\boldsymbol{w}_{\mathrm{F}}\right) = 0\,. \tag{38}$$

Moreover, keeping in mind the expression (32), we can remark that the difference between current and reference F-mass densities is finally given by the relation:

$$\varrho_{\mathrm{F}}(\mathbf{x},t) - \varrho^\star_{\mathrm{F0}}(\mathbf{X}) = \varepsilon\varrho^\star_{\mathrm{F1}}(\mathbf{X},t) - \varrho^\star_{\mathrm{F0}}(\mathbf{X})\;\mathrm{Div}\,\varepsilon\boldsymbol{u}_{\mathrm{S}}(\mathbf{X},t) + o\,(\varepsilon)\;. \tag{39}$$

## 4. Fluid-saturated poroelastic continua

In order to develop a macroscopic theory of saturated poroelastic continua (see e.g. Bowen [7, 5], Svendsen and Hutter [31], Klisch [20]), we shall enrich a self-consistent mathematical theory of binary solid-fluid mixtures by introducing the concept of volume fraction (Fillunger, see de Boer [10]; cf. Bluhm et al. [4], dell'Isola et al. [11]). In particular, we consider two independent scalar fields $\nu_\alpha$ (with $\alpha \in \{\mathrm{S,F}\}$ ), which represent the dimensionless ratio of the *macroscopic* mass density $\varrho_\alpha$ to the (constitutively prescribed) *microscopic* mass density $\hat{\varrho}_\alpha$ ,

$$\nu_\alpha\,(\mathbf{x},t) \;:=\; \frac{\varrho_\alpha\,(\mathbf{x},t)}{\hat{\varrho}_\alpha\left(\mathbf{X},\mathbf{E}_{\mathrm{S}}(\mathbf{X},t)\,,\varrho^\star_{\mathrm{F}}(\mathbf{X},t)\right)}\;,\quad \forall\,\mathbf{x} = \boldsymbol{\chi}_{\mathrm{S}}(\mathbf{X},t) \in \mathcal{B}(t)\,, \tag{40}$$

denoting by $\mathbf{E}_{\mathrm{S}}$ the Lagrangian strain tensor,

$$\mathbf{E}_{\mathrm{S}} = \frac{1}{2}\left(\mathbf{F}^{\mathsf{T}}_{\mathrm{S}}\mathbf{F}_{\mathrm{S}} - \mathbf{I}\right) = \varepsilon\,\mathrm{sym}\,(\mathrm{Grad}\;\boldsymbol{u}_{\mathrm{S}}) + o\,(\varepsilon)\;, \tag{41}$$

such that

$$\mathrm{tr}\;\mathbf{E}_{\mathrm{S}} \;\mathrel{\mathring{=}}\; \varepsilon\,\mathrm{Div}\,\boldsymbol{u}_{\mathrm{S}}\;. \tag{42}$$

A poroelastic solid infused with a compressible fluid is *saturated* if its solid skeleton is perfectly permeated by the fluid, i.e. if the *saturation constraint* is identically fulfilled:

$$\nu_{\mathrm{S}} + \nu_{\mathrm{F}} - 1 \;=\; 0\,. \tag{43}$$

This implies, by virtue of definitions (40) and (26), that

$$\varrho^\star_{\mathrm{S}}(\mathbf{X})\;\hat{\varrho}_{\mathrm{F}}(\mathbf{X},t) + \varrho^\star_{\mathrm{F}}(\mathbf{X},t)\;\hat{\varrho}_{\mathrm{S}}(\mathbf{X},t) = \hat{\varrho}_{\mathrm{S}}(\mathbf{X},t)\;\hat{\varrho}_{\mathrm{F}}(\mathbf{X},t)\;\det\mathbf{F}_{\mathrm{S}}(\mathbf{X},t)\,. \tag{44}$$

In order to linearize the algebraic equation (43), we consider a linear constitutive prescription for microscopic mass densities,

$$\hat{\varrho}_\alpha\big(\mathbf{X}, \mathbf{E}_{\mathrm{S}}(\mathbf{X},t)\,, \varrho^\star_{\mathrm{F}}(\mathbf{X},t)\big) \;\mathring{=}\; \hat{\varrho}_{\alpha 0}(\mathbf{X}) + \varepsilon \hat{\varrho}_{\alpha 1}(\mathbf{X},t)\ , \tag{45}$$

where†

$$\hat{\varrho}_{\alpha 1}(t) \;:=\; -\,\hat{\varrho}_{\alpha 0}\left(\lambda_{\alpha \mathrm{S}} \operatorname{Div} \boldsymbol{u}_{\mathrm{S}}(t) + \lambda_{\alpha \mathrm{F}}\,\frac{\varrho^\star_{\mathrm{F1}}(t)}{\hat{\varrho}_{\mathrm{F0}}}\right), \tag{46}$$

denoting by $\lambda_{\alpha\beta}$ ( $\alpha, \beta \in \{\mathrm{S}, \mathrm{F}\}$ ) a smooth time-independent scalar field on $\mathcal{B}$.

If we assume that the reference state is saturated,

$$\nu_{\mathrm{S0}} + \nu_{\mathrm{F0}} - 1 \;=\; 0 \qquad \Longleftrightarrow \qquad \varrho^\star_{\mathrm{S}}\hat{\varrho}_{\mathrm{F0}} + \varrho^\star_{\mathrm{F0}}\hat{\varrho}_{\mathrm{S0}} = \hat{\varrho}_{\mathrm{S0}}\hat{\varrho}_{\mathrm{F0}}\ , \tag{47}$$

with

$$\nu_{\mathrm{S0}} \;:=\; \frac{\varrho^\star_{\mathrm{S}}}{\hat{\varrho}_{\mathrm{S0}}} \qquad \text{and} \qquad \nu_{\mathrm{F0}} \;:=\; \frac{\varrho^\star_{\mathrm{F0}}}{\hat{\varrho}_{\mathrm{F0}}}\,, \tag{48}$$

then the algebraic equation (44) can be finally linearized, by virtue of expansions (35) and (45), resulting in:

$$\varrho^\star_{\mathrm{S}}\hat{\varrho}_{\mathrm{F1}}(t) + \varrho^\star_{\mathrm{F0}}\hat{\varrho}_{\mathrm{S1}}(t) + \hat{\varrho}_{\mathrm{S0}}\varrho^\star_{\mathrm{F1}}(t) = \hat{\varrho}_{\mathrm{S0}}\hat{\varrho}_{\mathrm{F1}}(t) + \hat{\varrho}_{\mathrm{F0}}\hat{\varrho}_{\mathrm{S1}}(t) + \hat{\varrho}_{\mathrm{S0}}\hat{\varrho}_{\mathrm{F0}}\operatorname{Div}\boldsymbol{u}_{\mathrm{S}}(t)\ , \tag{49}$$

and thus leading to the requirement:

$$\hat{\varrho}_{\mathrm{F0}}\,\beta_{\mathrm{S}}\operatorname{Div}\boldsymbol{u}_{\mathrm{S}}(t) + \beta_{\mathrm{F}}\,\varrho^\star_{\mathrm{F1}}(t) = 0\,, \tag{50}$$

with

$$\beta_{\mathrm{S}} := \nu_{\mathrm{F0}}\lambda_{\mathrm{FS}} + \nu_{\mathrm{S0}}\lambda_{\mathrm{SS}} - 1\,, \qquad \text{and} \qquad \beta_{\mathrm{F}} := \nu_{\mathrm{F0}}\lambda_{\mathrm{FF}} + \nu_{\mathrm{S0}}\lambda_{\mathrm{SF}} + 1\,. \tag{51}$$

It is worth mentioning (see Table 1) that if the coefficients $\lambda_{\alpha\beta}$ satisfy the two equations

$$\nu_{\mathrm{F0}}\lambda_{\mathrm{FS}} + \nu_{\mathrm{S0}}\lambda_{\mathrm{SS}} - 1 \;=\; 0 \qquad \text{and} \qquad \nu_{\mathrm{F0}}\lambda_{\mathrm{FF}} + \nu_{\mathrm{S0}}\lambda_{\mathrm{SF}} + 1 \;=\; 0 \tag{52}$$

for all $\mathbf{X} \in \mathcal{B}$ , then the constraint (50) is identically fulfilled by any possible fluid and solid motion. Otherwise, bearing in mind the expansion (35) and the equations (37)-(38), if the former of (52) is identically satisfied and the latter is not, then only F-motions characterized by the condition

$$\operatorname{Div}\big(\varrho^\star_{\mathrm{F0}}\boldsymbol{w}_{\mathrm{F}}(t)\big) \;=\; 0 \tag{53}$$

can be realized; analogously, if the latter of (52) is identically satisfied and the former is not, only isochoric S-motions are allowed. In the general case of $\beta_{\mathrm{S}} \neq 0$ and $\beta_{\mathrm{F}} \neq 0$ , the saturation constraint (50) just establishes a relation between the evolution of the fluid mass-density per unit reference volume of the mixture and the solid motion, that may also be rewritten in the alternative form:

$$\varrho^\star_{\mathrm{F}}(t) \;\mathring{=}\; \varrho^\star_{\mathrm{F0}}\left(1 - \frac{\beta_{\mathrm{S}}}{\nu_{\mathrm{F0}}\,\beta_{\mathrm{F}}}\operatorname{Div}\varepsilon\boldsymbol{u}_{\mathrm{S}}(t)\right)\,. \tag{54}$$

| | $\beta_F = 0$ | $\beta_F \neq 0$ |
|---|---|---|
| $\beta_S = 0$ | all S-motion<br>all F-motion | all S-motion<br>$\varrho^\star_{F1}(t) = 0$ |
| $\beta_S \neq 0$ | Div $\boldsymbol{u}_S(t) = 0$<br>all F-motion | $\hat{\varrho}_{F0}\, \beta_S$ Div $\boldsymbol{u}_S(t) + \beta_F\, \varrho^\star_{F1}(t) = 0$ |

Table 1. The saturation constraint.

Consistently, the expansions of volume fractions result in:

$$\nu_S(t) \ \mathring{=}\ \nu_{S0} + \varepsilon\nu_{S0}\,(\lambda_{SS} - 1)\,\mathrm{Div}\,\boldsymbol{u}_S(t) + \varepsilon\lambda_{SF}\,\nu_{S0}\,\frac{\varrho^\star_{F1}(t)}{\hat{\varrho}_{F0}} \tag{55}$$

$$\nu_F(t) \ \mathring{=}\ \nu_{F0} + \varepsilon\nu_{F0}\,(\lambda_{FS} - 1)\,\mathrm{Div}\,\boldsymbol{u}_S(t) + \varepsilon\,(\nu_{F0}\,\lambda_{FF} + 1)\,\frac{\varrho^\star_{F1}(t)}{\hat{\varrho}_{F0}}\,. \tag{56}$$

In order to highlight the role played by the coefficients $\lambda_{\alpha\beta}$ ( $\alpha, \beta \in \{\mathrm{S}, \mathrm{F}\}$ ) within the framework of the propounded constitutive theory, we observe in passing that, by virtue of definition (46) and hypothesis (45), it is possible to rewrite the expressions postulated for microscopic mass densities,

$$\hat{\varrho}_S(t) = \hat{\varrho}_{S0} - \lambda_{SS}\,\hat{\varrho}_{S0}\,\mathrm{Div}\,\varepsilon\boldsymbol{u}_S(t) - \lambda_{SF}\,\hat{\varrho}_{S0}\,\nu_{F0}\,\frac{\varepsilon\varrho^\star_{F1}(t)}{\varrho^\star_{F0}} \tag{57}$$

$$\hat{\varrho}_F(t) = \hat{\varrho}_{F0} - \lambda_{FS}\,\hat{\varrho}_{F0}\,\mathrm{Div}\,\varepsilon\boldsymbol{u}_S(t) - \lambda_{FF}\,\varepsilon\varrho^\star_{F1}(t)\ , \tag{58}$$

in a slightly different form, so as to emphasize their constitutive dependence on macroscopic (current) mass densities $\varrho_S(t)$ and $\varrho_F(t)$, namely:

$$\frac{\hat{\varrho}_S(t) - \hat{\varrho}_{S0}}{\hat{\varrho}_{S0}} = \lambda_{SS}\,\frac{\varrho_S(t) - \varrho^\star_S}{\varrho^\star_S} - \nu_{F0}\,\lambda_{SF}\,\frac{\varrho^\star_F(t) - \varrho^\star_{F0}}{\varrho^\star_{F0}} \tag{59}$$

$$\frac{\hat{\varrho}_F(t) - \hat{\varrho}_{F0}}{\hat{\varrho}_{F0}} = \lambda_{FS}\,\frac{\varrho_S(t) - \varrho^\star_S}{\varrho^\star_S} - \nu_{F0}\,\lambda_{FF}\,\frac{\varrho^\star_F(t) - \varrho^\star_{F0}}{\varrho^\star_{F0}}\,. \tag{60}$$

† For the sake of conciseness, from now on we shall drop the reference place $\mathbf{X}$ in any list of arguments. For instance, we shall write $\hat{\varrho}_{\alpha 0} + \varepsilon\hat{\varrho}_{\alpha 1}(t)$ in place of the right-hand side of first-order expansion (45).

## 5. Dynamics

### 5.1. Stress power

With the aim of describing local interactions exchanged by overlapped $\alpha$-points (figure 1) within the framework of a first-order gradient theory (Germain [17], Williams [36]), we assume the stress power per unit reference volume to be given by the expression:

$$\sum_{\alpha \in \{\mathrm{S},\mathrm{F}\}} \left( \mathrm{J}_{\mathrm{S}} \boldsymbol{\pi}_\alpha \cdot \mathbf{v}_\alpha + \mathbf{T}_\alpha \cdot \mathrm{Grad}\, \mathbf{v}_\alpha \right) , \tag{61}$$

where, respectively, $\mathbf{T}_\alpha$ denotes the partial Piola-Kirchhoff stress tensor associated with the $\alpha$-th constituent of the mixture (see, e.g. Bowen [7], Wilmanski [37, 38]), whereas $\pi_\alpha$ represents a general zeroth-order interaction (see e.g. dell'Isola et al. [11], Quiligotti et al. [27]).

According to the principle of material frame-indifference, the internal power density, expended on any rigid-body velocity field

$$\overset{\mathrm{R}}{\mathbf{v}}_{\mathrm{S}} (\mathbf{x}, t) = \overset{\mathrm{R}}{\mathbf{v}}_{\mathrm{F}} (\mathbf{x}, t) = \boldsymbol{\omega}(t) + \boldsymbol{\Omega}(t)(\mathbf{x} - \mathbf{x}_0) , \qquad \boldsymbol{\Omega}(t) \in \mathrm{Skw} , \tag{62}$$

needs to vanish for any choice of spatially uniform $\boldsymbol{\omega}(t)$ and $\boldsymbol{\Omega}(t) \in \mathrm{Skw}$. As a consequence, admissible constitutive assumptions have to satisfy the preliminary requirements:

$$\mathrm{skw}\left( (\mathbf{T}_{\mathrm{S}} + \mathbf{T}_{\mathrm{F}}) \mathbf{F}_{\mathrm{S}}^{\mathrm{T}} \right) = \mathbf{O} , \tag{63}$$

$$\boldsymbol{\pi}_{\mathrm{S}} + \boldsymbol{\pi}_{\mathrm{F}} = \mathbf{0} . \tag{64}$$

### 5.2. A constitutive theory of coupled interactions

Let us consider a material surface $\partial\mathcal{V}(\tau) \subset \mathcal{B}$, which migrates through the reference shape following the motion of the mixture as a whole. As S-points are fixed in $\mathcal{B}$ (namely, $\mathbf{w}_{\mathrm{S}} = \mathbf{0}$ in (29)), we shall assume that any material point of the mixture moves through the reference shape at the velocity

$$\mathbf{w} := \xi_{\mathrm{F}} \mathbf{w}_{\mathrm{F}} , \tag{65}$$

denoting by $\xi_{\mathrm{F}}$ the fluid-mass fraction,

$$\xi_{\mathrm{F}}(t) := \frac{\varrho^\star_{\mathrm{F}}(t)}{\varrho^\star_{\mathrm{S}} + \varrho^\star_{\mathrm{F}}(t)} . \tag{66}$$

Moreover, we assume that there exists an overall strain-energy density per unit reference volume of the mixture, $\mathfrak{W}(\mathbf{X}, t)$, such that the time derivative of the strain-energy content of $\mathcal{V}(\tau) \subset \mathcal{B}$,

$$\left\{ \frac{d}{d\tau} \int_{\mathcal{V}(\tau)} \mathfrak{W} \right\}_{\tau = t} = \int_{\mathcal{V}(t)} \left( \frac{\partial \mathfrak{W}}{\partial t} + \mathrm{Div}\,(\mathfrak{W}\, \mathbf{w}) \right) , \tag{67}$$

equals the stress power expended on the velocity pair $(\mathbf{v}_{\mathrm{S}}, \mathbf{v}_{\mathrm{F}})$,

$$\left\{ \frac{d}{d\tau} \int_{\mathcal{V}(\tau)} \mathfrak{W} \right\}_{\tau=t} = \sum_{\alpha \in \{\mathrm{S},\mathrm{F}\}} \int_{\mathcal{V}(t)} \left( \mathrm{J}_{\mathrm{S}}\, \boldsymbol{\pi}_\alpha \cdot \mathbf{v}_\alpha + \mathbf{T}_\alpha \cdot \mathrm{Grad}\, \mathbf{v}_\alpha \right) , \tag{68}$$

for any choice of $\mathcal{V}(t) \subset \mathcal{B}$. This leads to the local expression:

$$\frac{\partial \mathfrak{W}}{\partial t} + \mathrm{Div}\,(\mathfrak{W}\,\mathbf{w}) = \sum_{\alpha \in \{\mathrm{S},\mathrm{F}\}} \left( \mathrm{J}_{\mathrm{S}}\, \boldsymbol{\pi}_\alpha \cdot \mathbf{v}_\alpha + \mathbf{T}_\alpha \cdot \mathrm{Grad}\, \mathbf{v}_\alpha \right) , \tag{69}$$

whose right-hand side may be alternatively rewritten, by taking into account the relation (19) and the preliminary requirement (64), as

$$\mathbf{T} \cdot \mathrm{Grad}\, \mathbf{v}_{\mathrm{S}} + \boldsymbol{\tau} \cdot \mathbf{w}_{\mathrm{F}} + \boldsymbol{\mathcal{T}} \cdot \mathrm{Grad}\, \mathbf{w}_{\mathrm{F}} , \tag{70}$$

with

$$\mathbf{T} := \mathbf{T}_{\mathrm{S}} + \mathbf{T}_{\mathrm{F}} \tag{71}$$

$$\boldsymbol{\tau} := (\mathrm{Grad}\, \mathbf{F}_{\mathrm{S}})^{\mathsf{T}}\, \mathbf{T}_{\mathrm{F}} + \mathrm{J}_{\mathrm{S}}\, \mathbf{F}_{\mathrm{S}}^{\mathsf{T}} \boldsymbol{\pi}_{\mathrm{F}} \tag{72}$$

$$\boldsymbol{\mathcal{T}} := \mathbf{F}_{\mathrm{S}}^{\mathsf{T}} \mathbf{T}_{\mathrm{F}} . \tag{73}$$

## Unconstrained solid-fluid mixture

In order to investigate the physical meaning of local first-order interactions (71)-(73), let us consider an overall strain-energy density per unit reference volume of the mixture,

$$\mathfrak{W}(\mathbf{X},t) = \widehat{\mathfrak{W}}\left(\mathbf{X}, \mathbf{C}_{\mathrm{S}}(\mathbf{X},t), \varrho_{\mathrm{F}}^{\star}(\mathbf{X},t)\right) , \tag{74}$$

whose value, at any time $t$ and (reference) place $\mathbf{X} \in \mathcal{B}$, depends on:

(a) the reference place itself, i.e. the solid particle steadily associated with it (namely, $\mathfrak{X}_{\mathrm{S}} \in \mathfrak{B}_{\mathrm{S}}$ such that $\mathbf{X} = \mathcal{K}_{\mathrm{S}}(\mathfrak{X}_{\mathrm{S}}) \in \mathcal{B}$; see figure 1);

(b) the corresponding value of the right Cauchy-Green tensor ( $\mathbf{C}_{\mathrm{S}} = \mathbf{F}_{\mathrm{S}}^{\mathsf{T}} \mathbf{F}_{\mathrm{S}}$ );

(c) the value of the fluid mass density per unit reference volume of the corresponding fluid particle ( $\mathfrak{X}_{\mathrm{F}} \in \mathfrak{B}_{\mathrm{F}}$ ) which is placed, at the given time $t$, at the same current position occupied by the solid particle $\mathfrak{X}_{\mathrm{S}} \in \mathfrak{B}_{\mathrm{S}}$ (namely, $\mathbf{x} = \boldsymbol{\chi}_{\mathrm{S}}(\mathbf{X},t) = \boldsymbol{\chi}_{\mathrm{F}}(\mathfrak{X}_{\mathrm{F}},t) \in \mathcal{E}$ ).

With the aim of linearizing the equations that govern the nonlinear dynamics of the mixture, we may expand† in power series the strain-energy function in a neigh-borhood of a pre-stressed (saturated) reference state ( $\mathbf{C}_{\mathrm{S}} = \mathbf{I}$ , and $\varrho^{\star}_{\mathrm{F}} = \varrho^{\star}_{\mathrm{F0}}$ ),

$$\begin{aligned}\widehat{\mathfrak{W}}\left(\mathbf{X}, \mathbf{C}_{\mathrm{S}}, \varrho^{\star}_{\mathrm{F}}\right) \;\mathring{=}\; & \widehat{\mathfrak{W}}_o + \frac{1}{2}\,\mathbf{A}\cdot(\mathbf{C}_{\mathrm{S}} - \mathbf{I}) + a\left(\varrho^{\star}_{\mathrm{F}} - \varrho^{\star}_{\mathrm{F0}}\right) + \\ & + \frac{1}{8}\,\{\mathbb{B}\,(\mathbf{C}_{\mathrm{S}} - \mathbf{I})\}\cdot(\mathbf{C}_{\mathrm{S}} - \mathbf{I}) + \\ & + \frac{1}{2}\,b\left(\varrho^{\star}_{\mathrm{F}} - \varrho^{\star}_{\mathrm{F0}}\right)^2 + \frac{1}{2}\left(\varrho^{\star}_{\mathrm{F}} - \varrho^{\star}_{\mathrm{F0}}\right)\mathbf{B}\cdot(\mathbf{C}_{\mathrm{S}} - \mathbf{I})\ ,\end{aligned} \tag{75}$$

where $\widehat{\mathfrak{W}}_o = \widehat{\mathfrak{W}}\left(\mathbf{X}, \mathbf{I}, \varrho^{\star}_{\mathrm{F0}}\right)$ .

† For the sake of conciseness, we omit the arguments of $\mathbf{C}_{\mathrm{S}}$ and $\varrho_{\mathrm{F}}^{\star}$ .

As ( $\mathbf{C}_{\mathrm{S}} - \mathbf{I}$ ) is a symmetric tensor, only the symmetrical part of $\mathbf{A}, \mathbf{B}$ and $\mathbb{B}$ is responsible for any contribution to the strain-energy density. Thus, we shall assume these coefficients to be symmetrical. Moreover, as the strain-energy is supposed to be inhomogeneous, $a, b, \mathbf{A}, \mathbf{B}$ and $\mathbb{B}$ will generally depend on $\mathbf{X}$ , as well as $\widehat{\mathfrak{W}}_o$ . Recalling that

$$\frac{1}{2}\,(\mathbf{C}_{\mathrm{S}} - \mathbf{I}) \;\mathring{=}\; \varepsilon\,\mathrm{sym}\,(\mathrm{Grad}\,\boldsymbol{u}_{\mathrm{S}}) + \frac{1}{2}\,\varepsilon^2\,(\mathrm{Grad}\,\boldsymbol{u}_{\mathrm{S}})^{\mathrm{T}}\,\mathrm{Grad}\,\boldsymbol{u}_{\mathrm{S}}\ , \tag{76}$$

$$\varrho^{\star}_{\mathrm{F}} - \varrho^{\star}_{\mathrm{F0}} \;\mathring{=}\; \varepsilon\varrho^{\star}_{\mathrm{F1}}\ , \tag{77}$$

we finally obtain the expression:

$$\begin{aligned}\widehat{\mathfrak{W}} \;\mathring{=}\; & \widehat{\mathfrak{W}}_o + \mathbf{A}\cdot\varepsilon\,\mathrm{Grad}\,\boldsymbol{u}_{\mathrm{S}} + \varepsilon\,a\varrho^{\star}_{\mathrm{F1}} + \\ & + \frac{1}{2}\,\varepsilon^2\,\{(\mathrm{Grad}\,\boldsymbol{u}_{\mathrm{S}})\,\mathbf{A}\}\cdot\mathrm{Grad}\,\boldsymbol{u}_{\mathrm{S}} + \frac{1}{2}\,b\left(\varepsilon\varrho^{\star}_{\mathrm{F1}}\right)^2 + \\ & + \varepsilon^2\varrho^{\star}_{\mathrm{F1}}\mathbf{B}\cdot\mathrm{Grad}\,\boldsymbol{u}_{\mathrm{S}} + \frac{1}{2}\,\varepsilon^2\,\{\mathbb{B}\,(\mathrm{Grad}\,\boldsymbol{u}_{\mathrm{S}})\}\cdot\mathrm{Grad}\,\boldsymbol{u}_{\mathrm{S}}\ .\end{aligned} \tag{78}$$

As it will be useful later on, we notice, in passing, that the considered first-order approximation for the F-mass density per unit reference volume (35) leads the definition (66) of the F-mass fraction $\xi_{\mathrm{F}}(t)$ to yield:

$$\xi_{\mathrm{F}}(t) \;\mathring{=}\; \frac{\varrho^{\star}_{\mathrm{F0}} + \varepsilon\varrho^{\star}_{\mathrm{F1}}(t)}{\varrho^{\star}_{\mathrm{S}} + \varrho^{\star}_{\mathrm{F0}} + \varepsilon\varrho^{\star}_{\mathrm{F1}}(t)} \;\mathring{=}\; \xi_{\mathrm{F0}} + \varepsilon\xi_{\mathrm{F1}}(t)\ , \tag{79}$$

where $\xi_{\mathrm{F0}}$ and $\xi_{\mathrm{F1}}(t)$ are given, respectively, by the expressions:

$$\xi_{\mathrm{F0}} := \frac{\varrho^{\star}_{\mathrm{F0}}}{\varrho^{\star}_{\mathrm{S}} + \varrho^{\star}_{\mathrm{F0}}}\ , \qquad \text{and} \qquad \xi_{\mathrm{F1}}(t) := \frac{(1 - \xi_{\mathrm{F0}})\,\varrho^{\star}_{\mathrm{F1}}(t)}{\varrho^{\star}_{\mathrm{S}} + \varrho^{\star}_{\mathrm{F0}}}\ . \tag{80}$$

The analogous first-order approximation for the S-mass fraction $\xi_{\mathrm{S}}(t)$ reads:

$$\xi_{\mathrm{S}}(t) \;\mathring{=}\; \frac{\varrho^{\star}_{\mathrm{S}}}{\varrho^{\star}_{\mathrm{S}} + \varrho^{\star}_{\mathrm{F0}} + \varepsilon\varrho^{\star}_{\mathrm{F1}}(t)} \;\mathring{=}\; \xi_{\mathrm{S0}} + \varepsilon\xi_{\mathrm{S1}}(t)\ , \tag{81}$$

with

$$\xi_{\mathrm{S0}} := \frac{\varrho^{\star}_{\mathrm{S}}}{\varrho^{\star}_{\mathrm{S}} + \varrho^{\star}_{\mathrm{F0}}}\ , \qquad \text{and} \qquad \xi_{\mathrm{S1}}(t) := -\frac{\xi_{\mathrm{S0}}\,\varrho^{\star}_{\mathrm{F1}}(t)}{\varrho^{\star}_{\mathrm{S}} + \varrho^{\star}_{\mathrm{F0}}}\ . \tag{82}$$

Consistently, we can verify that the independent requirements:

$$\xi_{\rm F0} + \xi_{\rm S0} \;=\; 1\,, \tag{83}$$

$$\xi_{\rm F1}(t) + \xi_{\rm S1}(t) \;=\; 0\,, \tag{84}$$

are identically fulfilled at any time $t$.

With the aim of deducing from (69) an admissible set of constitutive prescriptions for the interactions (71)-(73), consistent with the assumption (78), we can now write down the resulting expression for the partial time derivative of the overall strain-energy density per unit reference volume,

$$\begin{aligned} \frac{\partial \mathfrak{W}}{\partial t} \;\mathrel{\overset{\circ}{=}}\; & \big\{[\mathbf{I} + \mathrm{Grad}\,(\varepsilon \boldsymbol{u}_{\rm S})]\,\mathbf{A} + \varepsilon \varrho^{\star}_{\rm F1}\mathbf{B} + \mathbb{B}\,\mathrm{Grad}\,(\varepsilon \boldsymbol{u}_{\rm S})\big\} \cdot \mathrm{Grad}\,(\varepsilon \boldsymbol{v}_{\rm S}) + \\ & - \varrho^{\star}_{\rm F0}\big\{a + b\,\varepsilon\varrho^{\star}_{\rm F1} + \mathbf{B}\cdot \mathrm{Grad}\,(\varepsilon \boldsymbol{u}_{\rm S})\big\}\,\mathbf{I}\cdot \mathrm{Grad}\,(\varepsilon \boldsymbol{w}_{\rm F}) + \\ & - \big\{a + b\,\varepsilon\varrho^{\star}_{\rm F1} + \mathbf{B}\cdot \mathrm{Grad}\,(\varepsilon \boldsymbol{u}_{\rm S})\big\}\,\mathrm{Grad}\,\varrho^{\star}_{\rm F0}\cdot \varepsilon \boldsymbol{w}_{\rm F}\,, \end{aligned} \tag{85}$$

keeping in mind equation (38) and definitions (13) and (20).

Moreover, recalling that $\mathbf{w} = \xi_{\rm F}\mathbf{w}_{\rm F}$ and considering a first-order approximation for the F-mass fraction $\xi_{\rm F}$ (79), we obtain:

$$\mathrm{Div}\,(\mathfrak{W}\,\mathbf{w}) = \mathbf{w}_{\rm F}\cdot \mathrm{Grad}\,(\xi_{\rm F}\mathfrak{W}) + \xi_{\rm F}\mathfrak{W}\,\mathbf{I}\cdot \mathrm{Grad}\,\mathbf{w}_{\rm F}\,, \tag{86}$$

where

$$\mathbf{w}_{\rm F}\cdot \mathrm{Grad}\,(\xi_{\rm F}\mathfrak{W}) \mathrel{\overset{\circ}{=}} \varepsilon \boldsymbol{w}_{\rm F}\cdot \left\{ \begin{aligned} & \mathrm{Grad}\,(\xi_{\rm F0}\widehat{\mathfrak{W}}_o) + \mathrm{Grad}\,(\varepsilon\xi_{\rm F1}\widehat{\mathfrak{W}}_o) + \\ & + \mathrm{Grad}\,\big(\xi_{\rm F0}\mathbf{A}\cdot \mathrm{Grad}\,(\varepsilon \boldsymbol{u}_{\rm S}) + \xi_{\rm F0}\,a\,\varepsilon\varrho^{\star}_{\rm F1}\big) \end{aligned} \right\}, \tag{87}$$

and

$$\xi_{\rm F}\mathfrak{W}\,\mathbf{I}\cdot \mathrm{Grad}\,\mathbf{w}_{\rm F} \mathrel{\overset{\circ}{=}} \mathrm{Grad}\,(\varepsilon \boldsymbol{w}_{\rm F})\cdot \left\{ \begin{aligned} & \xi_{\rm F0}\widehat{\mathfrak{W}}_o + \varepsilon\xi_{\rm F1}\widehat{\mathfrak{W}}_o + \xi_{\rm F0}\,a\,\varepsilon\varrho^{\star}_{\rm F1} + \\ & + \xi_{\rm F0}\mathbf{A}\cdot \mathrm{Grad}\,(\varepsilon \boldsymbol{u}_{\rm S}) \end{aligned} \right\}\mathbf{I}\,. \tag{88}$$

Henceforth, the coupled interactions (71)-(73) finally result in:

$$\mathbf{T}(t) \;\mathrel{\overset{\circ}{=}}\; \mathbf{T}_0 + \varepsilon\,\mathbf{T}_1(t) \tag{89}$$

$$\boldsymbol{\tau}(t) \;\mathrel{\overset{\circ}{=}}\; \boldsymbol{\tau}_0 + \varepsilon\,\boldsymbol{\tau}_1(t) \tag{90}$$

$$\boldsymbol{\mathcal{T}}(t) \;\mathrel{\overset{\circ}{=}}\; \boldsymbol{\mathcal{T}}_0 + \varepsilon\,\boldsymbol{\mathcal{T}}_1(t)\;, \tag{91}$$

with:

$$\mathbf{T}_0 \;=\; \mathbf{A} \tag{92}$$

$$\mathbf{T}_1(t) \;=\; \varrho^{\star}_{\rm F1}(t)\,\mathbf{B} \,+\, \{\mathrm{Grad}\,\boldsymbol{u}_{\rm S}(t)\}\,\mathbf{A} \,+\, \mathbb{B}\,\mathrm{Grad}\,\boldsymbol{u}_{\rm S}(t)\,, \tag{93}$$

$$\boldsymbol{\tau}_0 \;=\; -\,a\,\mathrm{Grad}\,\varrho^{\star}_{\rm F0} \,+\, \mathrm{Grad}\,(\xi_{\rm F0}\widehat{\mathfrak{W}}_o) \tag{94}$$

$$\begin{aligned} \boldsymbol{\tau}_1(t) \;=\; & -\big\{b\varrho^{\star}_{\rm F1}(t) + \mathbf{B}\cdot \mathrm{Grad}\,\boldsymbol{u}_{\rm S}(t)\big\}\,\mathrm{Grad}\,\varrho^{\star}_{\rm F0} \,+\, \mathrm{Grad}\,\Big\{\xi_{\rm F1}(t)\,\widehat{\mathfrak{W}}_o\Big\} \,+ \\ & + \mathrm{Grad}\,\big\{\xi_{\rm F0}\mathbf{A}\cdot \mathrm{Grad}\,\boldsymbol{u}_{\rm S}(t) + a\,\xi_{\rm F0}\,\varrho^{\star}_{\rm F1}(t)\big\}\;, \end{aligned} \tag{95}$$

$$\mathcal{T}_0 = \left(\xi_{F0}\widehat{\mathfrak{W}}_o - a\varrho^\star_{F0}\right)\mathbf{I} \tag{96}$$

$$\begin{aligned}\mathcal{T}_1(t) = &\left\{\xi_{F1}(t)\,\widehat{\mathfrak{W}}_o + a\,\xi_{F0}\,\varrho^\star_{F1}(t) + \xi_{F0}\mathbf{A}\cdot\mathrm{Grad}\,\boldsymbol{u}_S(t)\right\}\mathbf{I} + \\ &-\varrho^\star_{F0}\left\{b\varrho^\star_{F1}(t) + \mathbf{B}\cdot\mathrm{Grad}\,\boldsymbol{u}_S(t)\right\}\mathbf{I}\,.\end{aligned} \tag{97}$$

## Constrained solid-fluid mixture

As constraints are naturally associated with reactive actions, if the solid constituent is required to be perfectly permeated by the fluid, a *saturation pressure* $p\,(\mathbf{X},t)$ arises in the mixture so as to maintain each constituent in contact with the other one. Taking as granted that such a pressure does not expend power in any motion compatible with the constraint (43), we may regard the additional scalar field $p$ as a Lagrangian multiplier in the expression of the overall strain-energy density per unit reference volume,

$$\begin{aligned}\mathfrak{W}\,(\mathbf{X},t) &= \widehat{\mathfrak{W}}_c\left(\mathbf{X},\mathbf{C}_S(\mathbf{X},t),\varrho^\star_F(\mathbf{X},t),p\,(\mathbf{X},t)\right) = \\ &= \widehat{\mathfrak{W}}\left(\mathbf{X},\mathbf{C}_S(\mathbf{X},t),\varrho^\star_F(\mathbf{X},t)\right) + p\,(\mathbf{X},t)\left\{\nu_S\,(\mathbf{X},t)+\nu_F\,(\mathbf{X},t)-1\right\}\,,\end{aligned} \tag{98}$$

such that, expanding in power series the strain-energy function in a neighborhood of a pre-stressed (saturated) reference state ( $\mathbf{C}_S=\mathbf{I}$, $\varrho^\star_F=\varrho^\star_{F0}$ , $p=p_0$ ), and bearing in mind the expressions (50) and (51), the second-order approximation of the last term of the strain-energy function (98) results[†] in:

$$p\,(t)\left\{\nu_S(t)+\nu_F(t)-1\right\} \mathrel{\mathring{=}} \left\{p_0+\varepsilon p_1(t)\right\}\left\{\beta_S\,\mathrm{Div}\,\varepsilon\boldsymbol{u}_S(t) + \beta_F\,\frac{\varepsilon\varrho^\star_{F1}(t)}{\hat{\varrho}_{F0}}\right\}, \tag{99}$$

so as to get (compare with (78)), in the end,

$$\begin{aligned}\widehat{\mathfrak{W}} \mathrel{\mathring{=}}\ &\widehat{\mathfrak{W}}_o + \tilde{\mathbf{A}}\cdot\varepsilon\,\mathrm{Grad}\,\boldsymbol{u}_S + \varepsilon\,\tilde{a}\varrho^\star_{F1} + \frac{1}{2}\,b\left(\varepsilon\varrho^\star_{F1}\right)^2 + \\ &+\frac{1}{2}\,\varepsilon^2\left\{(\mathrm{Grad}\,\boldsymbol{u}_S)\,\mathbf{A}\right\}\cdot\mathrm{Grad}\,\boldsymbol{u}_S + \varepsilon^2\varrho^\star_{F1}\mathbf{B}\cdot\mathrm{Grad}\,\boldsymbol{u}_S + \\ &+\frac{1}{2}\,\varepsilon^2\left\{\mathbb{B}\,\mathrm{Grad}\,\boldsymbol{u}_S\right\}\cdot\mathrm{Grad}\,\boldsymbol{u}_S\,,\end{aligned} \tag{100}$$

with

$$\tilde{\mathbf{A}}\,(t) := \mathbf{A} + p_0\beta_S\,\mathbf{I} + \varepsilon\,p_1(t)\,\beta_S\,\mathbf{I}\,, \tag{101}$$

$$\tilde{a}\,(t) := a + \beta_F\,\frac{p_0}{\hat{\varrho}_{F0}} + \varepsilon\,\beta_F\,\frac{p_1(t)}{\hat{\varrho}_{F0}}\,. \tag{102}$$

Following the same procedure as that outlined in the former section, it is possible to deduce from (100) a suitable expression for the generalized coupled forces

[†] For the sake of conciseness, we shall drop the reference place $\mathbf{X}$ in any list of arguments.

(71)-(73), so as to describe a first-order interaction between the fluid-saturated poroelastic solid and the filling fluid, namely (compare with (92)-(97)):

$$\mathbf{T}_{0} = \mathbf{A} + p_0 \beta_{S}\, \mathbf{I} \tag{103}$$

$$\mathbf{T}_{1}(t) = \varrho^{\star}_{F1}(t)\, \mathbf{B} + \{\mathrm{Grad}\, \boldsymbol{u}_{S}(t)\}\, \mathbf{A} + + \mathbb{B}\, \mathrm{Grad}\, \boldsymbol{u}_{S}(t) + p_1(t)\, \beta_{S}\, \mathbf{I}, \tag{104}$$

$$\boldsymbol{\tau}_{0} = -\left(a + \beta_{F}\frac{p_0}{\hat{\varrho}_{F0}}\right) \mathrm{Grad}\, \varrho^{\star}_{F0} + \mathrm{Grad}\,(\xi_{F0}\widehat{\mathfrak{W}}_o) \tag{105}$$

$$\begin{aligned}\boldsymbol{\tau}_{1}(t) = & -\left\{b\varrho^{\star}_{F1}(t) + \mathbf{B}\cdot \mathrm{Grad}\, \boldsymbol{u}_{S}(t) + \beta_{F}\frac{p_1(t)}{\hat{\varrho}_{F0}}\right\} \mathrm{Grad}\, \varrho^{\star}_{F0} + \\ & + \mathrm{Grad}\,\{\xi_{F0}\,(\mathbf{A} + p_0\beta_{S}\mathbf{I})\cdot \mathrm{Grad}\, \boldsymbol{u}_{S}(t)\} + \\ & + \mathrm{Grad}\left\{\xi_{F0}\left(a + \beta_{F}\frac{p_0}{\hat{\varrho}_{F0}}\right)\varrho^{\star}_{F1}(t) + \xi_{F1}(t)\,\widehat{\mathfrak{W}}_o\right\},\end{aligned} \tag{106}$$

$$\boldsymbol{\mathcal{T}}_{0} = \left(\xi_{F0}\widehat{\mathfrak{W}}_o - a\,\varrho^{\star}_{F0} - \nu_{F0}\beta_{F}p_0\right)\mathbf{I} \tag{107}$$

$$\begin{aligned}\boldsymbol{\mathcal{T}}_{1}(t) = & \left\{\xi_{F1}(t)\,\widehat{\mathfrak{W}}_o + \xi_{F0}\left(a + \beta_{F}\frac{p_0}{\hat{\varrho}_{F0}}\right)\varrho^{\star}_{F1}(t)\right\}\mathbf{I} + \\ & + \{\xi_{F0}\,(\mathbf{A} + p_0\beta_{S}\mathbf{I})\cdot \mathrm{Grad}\, \boldsymbol{u}_{S}(t)\}\,\mathbf{I} + \\ & - \varrho^{\star}_{F0}\left\{b\varrho^{\star}_{F1}(t) + \mathbf{B}\cdot \mathrm{Grad}\, \boldsymbol{u}_{S}(t) + \beta_{F}\frac{p_1(t)}{\hat{\varrho}_{F0}}\right\}\mathbf{I}.\end{aligned} \tag{108}$$

### 5.3. Kinetic energy

We assume (cf. Biot [3], de Boer [10], Edelman and Wilmanski [14] ) that the kinetic energy density per unit reference volume of the mixture is given by the sum

$$\sum_{\alpha\in\{S,F\}} \frac{1}{2}\,\varrho^{\star}_{\alpha}\mathbf{v}_{\alpha}\cdot\mathbf{v}_{\alpha} = \frac{1}{2}\,\varrho^{\star}\mathbf{v}\cdot\mathbf{v} - \frac{1}{2}\,\varrho^{\star}\,(\mathbf{d}_{S}\cdot\mathbf{d}_{F}) \tag{109}$$

where $\varrho^{\star}$ and $\mathbf{v}$ are, respectively, the overall mass density per unit reference volume and the mean velocity of mixture particles,

$$\varrho^{\star} := \varrho^{\star}_{S} + \varrho^{\star}_{F}, \tag{110}$$

$$\mathbf{v} := \xi_{S}\,\mathbf{v}_{S} + \xi_{F}\,\mathbf{v}_{F} = \mathbf{v}_{S} + \xi_{F}\,\mathbf{F}_{S}\mathbf{w}_{F}, \tag{111}$$

whereas $\mathbf{d}_{\alpha}$ denotes the diffusion velocity of the $\alpha$-th constituent,

$$\mathbf{d}_{S} := \mathbf{v}_{S} - \mathbf{v} = \xi_{F}\,(\mathbf{v}_{S} - \mathbf{v}_{F}), \tag{112}$$

$$\mathbf{d}_{F} := \mathbf{v}_{F} - \mathbf{v} = \xi_{S}\,(\mathbf{v}_{F} - \mathbf{v}_{S}). \tag{113}$$

The time derivative of the kinetic energy associated with any smooth region of the reference shape, $\mathcal{V}(\tau) \subset \mathcal{B}$, enveloped by a migrating surface which follows the motion of the mixture as a whole, is required to equal, at any time $t$, the integral over $\mathcal{V}(t)$ of the power expended by inertial forces,

$$\sum_{\alpha\in\{\mathrm{S},\mathrm{F}\}} \left\{ \frac{d}{d\tau} \int_{\mathcal{V}(\tau)} \frac{1}{2}\, \varrho^{\star}_{\alpha} \mathbf{v}_{\alpha} \cdot \mathbf{v}_{\alpha} \right\}_{\tau=t} = \int_{\mathcal{V}(t)} (\mathbf{v}_{\mathrm{S}} \cdot \mathbf{f}_{\mathrm{S}} + \mathbf{w}_{\mathrm{F}} \cdot \mathbf{f}_{\mathrm{F}})\,, \tag{114}$$

whose local expression is given† by

$$\mathbf{f}_{\mathrm{S}} := \varrho^{\star} \frac{\mathcal{D}\mathbf{v}}{\mathcal{D}t}\,, \tag{115}$$

$$\mathbf{f}_{\mathrm{F}} := \frac{1}{2}\, \varrho^{\star}\, \mathbf{F}^{\mathsf{T}}_{\mathrm{S}} \left( \frac{\mathcal{D}\mathbf{v}}{\mathcal{D}t} - \xi_{\mathrm{S}} \frac{\mathcal{D}\mathbf{v}_{\mathrm{S}}}{\mathcal{D}t} + \xi_{\mathrm{F}} \frac{\mathcal{D}\mathbf{v}_{\mathrm{F}}}{\mathcal{D}t} \right). \tag{116}$$

In order to linearize the equations that govern the nonlinear dynamics of the mixture, we notice, by taking into account a first-order approximations of $\varrho^{\star}$, $\mathbf{F}^{\mathsf{T}}_{\mathrm{S}}$, $\mathbf{v}$, $\mathbf{v}_{\mathrm{S}}$, and $\mathbf{v}_{\mathrm{F}}$ (see sections 2 and 3), that linearized expressions for (115) and (116) result in:

$$\mathbf{f}_{\mathrm{F}} \mathrel{\mathring{=}} \mathbf{f}_{\mathrm{F}0} + \varepsilon\, \mathbf{f}_{\mathrm{F}1} = \varepsilon \left( \varrho^{\star}_{\mathrm{F}0} \frac{\partial^2 \boldsymbol{u}_{\mathrm{S}}}{\partial t^2} + \varrho^{\star}_{\mathrm{F}0} \frac{\partial \boldsymbol{w}_{\mathrm{F}}}{\partial t} \right), \tag{117}$$

$$\mathbf{f}_{\mathrm{S}} \mathrel{\mathring{=}} \mathbf{f}_{\mathrm{S}0} + \varepsilon\, \mathbf{f}_{\mathrm{S}1} = \varepsilon \left( (\varrho^{\star}_{\mathrm{S}} + \varrho^{\star}_{\mathrm{F}0}) \frac{\partial^2 \boldsymbol{u}_{\mathrm{S}}}{\partial t^2} + \varrho^{\star}_{\mathrm{F}0} \frac{\partial \boldsymbol{w}_{\mathrm{F}}}{\partial t} \right), \tag{118}$$

with $\mathbf{f}_{\mathrm{F}0} = \mathbf{f}_{\mathrm{S}0} = \mathbf{0}$.

## 5.4. Governing equations

The required set of balance equations and boundary conditions, that governs the nonlinear dynamics of the mixture, can be straightforwardly deduced from a general statement of the principle of virtual power (cf. Maugin [22]).

As the thorny question of splitting the overall applied boundary traction among the constituents still stands as one of the greatest challenges that have to be faced up in order to put mixture theories into use (see e.g. Rajagopal and Tao [28], Reynolds and Humphrey [29]), it is worth emphasizing that, within the framework of variational principles, boundary conditions are derived as a result of the theory, as well as governing equations.

† The differential operator

$$\frac{\mathcal{D}(\star)}{\mathcal{D}t} := \frac{\partial(\star)}{\partial t} + [\mathrm{Grad}\,(\star)]\,\mathbf{w}$$

denotes the material derivative of a given vector field $(\star)$, defined on the reference configuration $\mathcal{B}$, following the motion of the mixture as a single body.

If the total power expended on any conceivable pair of smooth *test* velocity fields $\hat{\mathbf{v}}_s$ and $\hat{\mathbf{w}}_f$ (in absence of applied body forces) is required to vanish,

$$\int_{\partial\mathcal{B}} \left(\mathbf{t}\cdot\hat{\mathbf{v}}_s + \xi_{\mathrm{F}}\mathbf{F}_{\mathrm{S}}^{\mathsf{T}}\mathbf{t}\cdot\hat{\mathbf{w}}_f\right) - \int_{\mathcal{B}} \left(\hat{\mathbf{v}}_s\cdot\mathbf{f}_{\mathrm{S}} + \hat{\mathbf{w}}_f\cdot\mathbf{f}_{\mathrm{F}}\right) + \\ - \int_{\mathcal{B}} \left(\mathbf{T}\cdot\operatorname{Grad}\hat{\mathbf{v}}_s + \boldsymbol{\tau}\cdot\hat{\mathbf{w}}_f + \boldsymbol{\mathcal{T}}\cdot\operatorname{Grad}\hat{\mathbf{w}}_f\right) = 0\,, \tag{119}$$

the resulting set of nonlinear balance equations and boundary conditions is:

$$\left.\begin{aligned} \operatorname{Div}\mathbf{T} &= \mathbf{f}_{\mathrm{S}} \\ \operatorname{Div}\boldsymbol{\mathcal{T}} - \boldsymbol{\tau} &= \mathbf{f}_{\mathrm{F}} \end{aligned}\right\} \quad \text{on} \quad \mathcal{B}\subset\mathcal{E}$$
$$\left.\begin{aligned} \mathbf{T}\mathbf{n} &= \mathbf{t} \\ \boldsymbol{\mathcal{T}}\mathbf{n} &= \xi_{\mathrm{F}}\mathbf{F}_{\mathrm{S}}^{\mathsf{T}}\mathbf{t} \end{aligned}\right\} \quad \text{on} \quad \partial\mathcal{B}\subset\mathcal{E}\,, \tag{120}$$

where $\mathbf{t}$ represents the overall applied boundary traction.

These equations may be linearized by taking into account first-order expansions of all vector and tensor fields involved. In particular, recalling that $\mathbf{f}_{\mathrm{S}0} = \mathbf{f}_{\mathrm{F}0} = \mathbf{0}$ (section 5.3), it is possible to deduce the set of balance equations and boundary conditions that characterize the reference state,

$$\left.\begin{aligned} \operatorname{Div}\mathbf{T}_0 &= \mathbf{0} \\ \operatorname{Div}\boldsymbol{\mathcal{T}}_0 - \boldsymbol{\tau}_0 &= \mathbf{0} \end{aligned}\right\} \quad \text{on} \quad \mathcal{B}\subset\mathcal{E}$$
$$\left.\begin{aligned} \mathbf{T}_0\mathbf{n} &= \mathbf{t}_0 \\ \boldsymbol{\mathcal{T}}_0\mathbf{n} &= \xi_{\mathrm{F}0}\mathbf{t}_0 \end{aligned}\right\} \quad \text{on} \quad \partial\mathcal{B}\subset\mathcal{E}\,, \tag{121}$$

whereas the perturbed state is governed by the equations:

$$\left.\begin{aligned} \operatorname{Div}\mathbf{T}_1 &= \left(\varrho_{\mathrm{S}}^{\star} + \varrho_{\mathrm{F}0}^{\star}\right)\frac{\partial^2\boldsymbol{u}_{\mathrm{S}}}{\partial t^2} + \varrho_{\mathrm{F}0}^{\star}\frac{\partial\boldsymbol{w}_{\mathrm{F}}}{\partial t} \\ \operatorname{Div}\boldsymbol{\mathcal{T}}_1 - \boldsymbol{\tau}_1 &= \varrho_{\mathrm{F}0}^{\star}\frac{\partial^2\boldsymbol{u}_{\mathrm{S}}}{\partial t^2} + \varrho_{\mathrm{F}0}^{\star}\frac{\partial\boldsymbol{w}_{\mathrm{F}}}{\partial t} \end{aligned}\right\} \quad \text{on} \quad \mathcal{B}\subset\mathcal{E}$$
$$\left.\begin{aligned} \mathbf{T}_1\mathbf{n} &= \mathbf{t}_1 \\ \boldsymbol{\mathcal{T}}_1\mathbf{n} &= \xi_{\mathrm{F}0}\mathbf{t}_1 + \xi_{\mathrm{F}0}\left(\operatorname{Grad}\boldsymbol{u}_{\mathrm{S}}\right)^{\mathsf{T}}\mathbf{t}_0 + \xi_{\mathrm{F}1}\mathbf{t}_0 \end{aligned}\right\} \quad \text{on} \quad \partial\mathcal{B}\subset\mathcal{E}\,. \tag{122}$$

### 5.5. Elastic waves in unbounded solid-fluid mixtures

A plane harmonic wave (see e.g. Achenbach [1], Graff [18]), propagating with phase velocity $c$ in a direction defined by the unit vector $\mathbf{q}$, can be generally represented by the real (or the imaginary) part of a complex function,

$$\boldsymbol{\psi}\left(\mathbf{X},t\right) = \tilde{\boldsymbol{\psi}}\left(\omega\right)\, e^{ik(\mathbf{X}\cdot\,\mathbf{q}-ct)}\,, \tag{123}$$

whose amplitude $\tilde{\boldsymbol{\psi}}(\omega)$ is independent of $(\mathbf{X}, t)$.

By definition, the characteristic wavenumber $k$ is related to the circular (or radial) frequency $\omega$ by the algebraic relation $\omega = kc$, while the wavelength $\Lambda$ is given by tha ratio $\Lambda = 2\pi/k$.

With the aim of investigating the propagation of harmonic displacement waves in unbounded media, we shall assume homogeneity and isotropy of constitutive prescriptions,

$$\varrho^{\star}_{\mathrm{F1}}(t)\,\mathbf{B}\cdot\mathrm{Grad}\,\boldsymbol{u}_{\mathrm{S}}(t) \;:=\; \varrho^{\star}_{\mathrm{F1}}(t)\;\beta\,\mathbf{I}\cdot\mathrm{Grad}\,\boldsymbol{u}_{\mathrm{S}}(t) = \varrho^{\star}_{\mathrm{F1}}(t)\,\beta\,\mathrm{Div}\,\boldsymbol{u}_{\mathrm{S}}(t) \tag{124}$$

$$\mathbb{B}\,\mathrm{Grad}\,\boldsymbol{u}_{\mathrm{S}}(t) \;:=\; \lambda_{\mathrm{S}}\mathrm{Div}\,\boldsymbol{u}_{\mathrm{S}}(t)\;\mathbf{I} + \mu_{\mathrm{S}}\mathrm{Grad}\,\boldsymbol{u}_{\mathrm{S}}(t) + \mu_{\mathrm{S}}\mathrm{Grad}^{\mathsf{T}}\boldsymbol{u}_{\mathrm{S}}(t) \tag{125}$$

$$\mathbf{A} \;:=\; \alpha\,\mathbf{I}, \tag{126}$$

where $\alpha$, $\beta$, $\mu_{\mathrm{S}}$ and $\lambda_{\mathrm{S}}$ are assumed to be constant in the chosen reference state, furthermore characterized by a uniform (macroscopic) fluid-mass density ($\mathrm{Grad}\,\varrho^{\star}_{\mathrm{F0}} = \mathbf{0}$), and a constant (macroscopic) fluid compressibility (namely, $\mathrm{Grad}\,b = \mathbf{0}$). Moreover, for the sake of simplicity, we focus our attention on uniform (microscopic) mass densities per unit reference volume ($\mathrm{Grad}\,\hat{\varrho}_{\mathrm{S0}} = \mathbf{0}$, $\mathrm{Grad}\,\hat{\varrho}_{\mathrm{F0}} = \mathbf{0}$), so as to deal with a constant reference porosity ($\mathrm{Grad}\,\nu_{\mathrm{F0}} = \mathbf{0}$).

### Unconstrained solid-fluid mixture

At first, we investigate the problem of the propagation of harmonic plane waves in unbounded and *unconstrained* solid-fluid mixtures, whose linearized dynamics is governed by the fluid-mass conservation law (38) and the set of field equations $(122)_1$. Recalling that, by assumption,

$$\mathrm{Grad}\,\varrho^{\star}_{\mathrm{F0}} \;=\; \mathbf{0} \tag{127}$$

and

$$\begin{aligned}\mathbf{T}_1(t) \;&=\; \left\{\beta\varrho^{\star}_{\mathrm{F1}}(t) + \lambda_{\mathrm{S}}\mathrm{Div}\,\boldsymbol{u}_{\mathrm{S}}(t)\right\}\mathbf{I} + \\ &\quad + 2\mu_{\mathrm{S}}\,\mathrm{sym}\,(\mathrm{Grad}\,\boldsymbol{u}_{\mathrm{S}}(t)) + \alpha\,\mathrm{Grad}\,\boldsymbol{u}_{\mathrm{S}}(t)\end{aligned} \tag{128}$$

$$\boldsymbol{\tau}_1(t) \;=\; \mathrm{Grad}\left\{\xi_{\mathrm{F1}}(t)\,\widehat{\mathfrak{W}}_o + \xi_{\mathrm{F0}}\mathbf{A}\cdot\mathrm{Grad}\,\boldsymbol{u}_{\mathrm{S}}(t) + a\,\xi_{\mathrm{F0}}\varrho^{\star}_{\mathrm{F1}}(t)\right\} \tag{129}$$

$$\begin{aligned}\boldsymbol{\mathcal{T}}_1(t) \;&=\; \left\{\xi_{\mathrm{F1}}(t)\,\widehat{\mathfrak{W}}_o \,+\, a\,\xi_{\mathrm{F0}}\varrho^{\star}_{\mathrm{F1}}(t) \,+\, \xi_{\mathrm{F0}}\mathbf{A}\cdot\mathrm{Grad}\,\boldsymbol{u}_{\mathrm{S}}(t)\right\}\mathbf{I} + \\ &\quad - \varrho^{\star}_{\mathrm{F0}}\left\{b\varrho^{\star}_{\mathrm{F1}}(t) + \beta\,\mathrm{Div}\,\boldsymbol{u}_{\mathrm{S}}(t)\right\}\mathbf{I},\end{aligned} \tag{130}$$

it is possible to formulate the problem in terms of the field triplet $\{\boldsymbol{u}_{\rm S}, \boldsymbol{w}_{\rm F}, \varrho^\star_{\rm F1}\}$,

$$\frac{\partial \varrho^\star_{\rm F1}}{\partial t} + \varrho^\star_{\rm F0} \operatorname{Div} \boldsymbol{w}_{\rm F} = 0 \qquad (131)$$

$$\begin{aligned}\left(\varrho^\star_{\rm S} + \varrho^\star_{\rm F0}\right) \frac{\partial^2 \boldsymbol{u}_{\rm S}}{\partial t^2} + \varrho^\star_{\rm F0} \frac{\partial \boldsymbol{w}_{\rm F}}{\partial t} - (\lambda_{\rm S} + \mu_{\rm S}) \operatorname{Grad}(\operatorname{Div} \boldsymbol{u}_{\rm S}) + & \\ - (\mu_{\rm S} + \alpha) \operatorname{Div}(\operatorname{Grad} \boldsymbol{u}_{\rm S}) - \beta \operatorname{Grad} \varrho^\star_{\rm F1} &= \mathbf{0} \end{aligned} \qquad (132)$$

$$\varrho^\star_{\rm F0} \frac{\partial^2 \boldsymbol{u}_{\rm S}}{\partial t^2} + \varrho^\star_{\rm F0} \frac{\partial \boldsymbol{w}_{\rm F}}{\partial t} + \varrho^\star_{\rm F0} \beta \operatorname{Grad}(\operatorname{Div} \boldsymbol{u}_{\rm S}) + b \varrho^\star_{\rm F0} \operatorname{Grad} \varrho^\star_{\rm F1} = \mathbf{0}, \qquad (133)$$

whereas the (unperturbed) reference state satisfies the set of linear field equations deducible from $(121)_1$, namely

$$\begin{cases} \operatorname{Div} \mathbf{T}_0 = \operatorname{Div} \mathbf{A} = \operatorname{Grad} \alpha = \mathbf{0} \\ \operatorname{Div} \boldsymbol{\mathcal{T}}_0 - \boldsymbol{\tau}_0 = -\varrho^\star_{\rm F0} \operatorname{Grad} a = \mathbf{0}. \end{cases} \qquad (134)$$

Looking for steady-state solutions in the form:

$$\boldsymbol{u}_{\rm S}(\mathbf{X}, t) = \tilde{\boldsymbol{u}}\, e^{ik(\mathbf{X}\cdot\,\mathbf{q} - ct)} \qquad (135)$$

$$\boldsymbol{w}_{\rm F}(\mathbf{X}, t) = -ikc\, \tilde{\boldsymbol{w}}\, e^{ik(\mathbf{X}\cdot\,\mathbf{q} - ct)} \qquad (136)$$

$$\varrho^\star_{\rm F1}(\mathbf{X}, t) = \tilde{\varrho}\, e^{ik(\mathbf{X}\cdot\,\mathbf{q} - ct)}, \qquad (137)$$

we find out, by virtue of equation (131), a relation between $\tilde{\varrho}$ and $\tilde{\boldsymbol{w}}$,

$$\tilde{\varrho} = -ik\varrho^\star_{\rm F0}\, (\tilde{\boldsymbol{w}} \cdot \mathbf{q}), \qquad (138)$$

which may be used to uncouple the subset of equations (132)-(133) from the fluid mass-conservation law (131). Henceforth, we can at first focus our attention on the resulting subset of (algebraic) equations (Meirovitch [24, 25]) that governs the steady-state linearized dynamics of the mixture,

$$\left(\mathbf{K} - c^2 \mathbf{M}\right) \boldsymbol{X} = \mathbf{O}, \qquad (139)$$

where $\mathbf{M}$, $\mathbf{K}$ and $\boldsymbol{X}$ denote, respectively:

$$\{\mathbf{M}\} := \begin{Bmatrix} \left(\varrho^\star_{\rm S} + \varrho^\star_{\rm F0}\right) \mathbf{I} & \varrho^\star_{\rm F0} \mathbf{I} \\ \varrho^\star_{\rm F0} \mathbf{I} & \varrho^\star_{\rm F0} \mathbf{I} \end{Bmatrix} \qquad (140)$$

$$\{\mathbf{K}\} := \begin{Bmatrix} \varrho^\star_{\rm S} c^2_{\rm TS} \mathbf{I} + \varrho^\star_{\rm S} \left(c^2_{\rm LS} - c^2_{\rm TS}\right) (\mathbf{q} \otimes \mathbf{q}) & -\beta \varrho^\star_{\rm F0} (\mathbf{q} \otimes \mathbf{q}) \\ -\beta \varrho^\star_{\rm F0} (\mathbf{q} \otimes \mathbf{q}) & \varrho^\star_{\rm F0} c^2_{\rm LF} (\mathbf{q} \otimes \mathbf{q}) \end{Bmatrix} \qquad (141)$$

$$\{\boldsymbol{X}\} := \begin{Bmatrix} \tilde{\boldsymbol{u}} \\ \tilde{\boldsymbol{w}} \end{Bmatrix}, \qquad (142)$$

with

$$c^2_{\rm LS} = \frac{\lambda_{\rm S} + 2\mu_{\rm S} + \alpha}{\varrho^\star_{\rm S}}, \quad c^2_{\rm TS} = \frac{\mu_{\rm S} + \alpha}{\varrho^\star_{\rm S}}, \quad \text{and} \quad c^2_{\rm LF} = \varrho^\star_{\rm F0} b\,. \qquad (143)$$

We notice in passing that the mass matrix $\mathbf{M}$ is symmetrical and positive definite for any referential value of mass fractions belonging to the open set $(0,1)$ of the real axis. Its eigenvalues (figure 2)

$$\frac{1}{2}\,\varrho_0^\star\left(2\,\xi_{\mathrm{F0}}+\xi_{\mathrm{S0}}\pm\sqrt{4\,\xi_{\mathrm{F0}}^2+\xi_{\mathrm{S0}}^2}\right)=:\varrho_0^\star\,h^{\pm}(\xi_{\mathrm{S0}})\,, \tag{144}$$

characterized by a triple multiplicity, are associated with the following set of linearly independent eigenvectors,

$$\left\{\begin{array}{c} 2\,\xi_{\mathrm{F0}}\,\boldsymbol{e}_j \\ g^{\pm}(\xi_{\mathrm{S0}})\,\boldsymbol{e}_j \end{array}\right\};\quad j\in\{1,2,3\}\,, \tag{145}$$

where $\boldsymbol{e}_i\cdot\boldsymbol{e}_j=\delta_{ij}$, and (figure 2)

$$g^{\pm}(\xi_{\mathrm{S0}}):=\pm\sqrt{4\,\xi_{\mathrm{F0}}^2+\xi_{\mathrm{S0}}^2}-\xi_{\mathrm{S0}}\,. \tag{146}$$

Moreover, it can be shown that $\det\mathbf{M}=\left(\varrho_{\mathrm{F0}}^\star\varrho_{\mathrm{S}}^\star\right)^3$ (figure 3).

The stiffness matrix $\mathbf{K}$ is also symmetrical. Furthermore, it is positive semidefinite† if the (macroscopic) coupling coefficient $\beta$, introduced in (75) by means of assumption (124), meets the requirement:

$$|\,\beta\,|\leq\beta_{max}\,,\qquad \beta_{max}:=\sqrt{\frac{\varrho_{\mathrm{S}}^\star}{\varrho_{\mathrm{F0}}^\star}\;c_{\mathrm{LS}}^2c_{\mathrm{LF}}^2}=\sqrt{b\,(\lambda_{\mathrm{S}}+2\mu_{\mathrm{S}}+\alpha)}\,. \tag{147}$$

General features of both eigenvectors and eigenvalues of $\mathbf{K}$ are summarized in tables 2, 3 and 4.

| | $\tilde{\boldsymbol{u}}_k=\mathbf{0}$ | $\tilde{\boldsymbol{u}}_k=g_1^{\pm}(\beta)\,\mathbf{q}$ | $\tilde{\boldsymbol{u}}_k\cdot\mathbf{q}=0$ |
|---|---|---|---|
| $\tilde{\boldsymbol{w}}_k=\mathbf{0}$ | — | — | $\varrho_{\mathrm{S}}^\star c_{\mathrm{TS}}^2$ |
| $\tilde{\boldsymbol{w}}_k=g_2^{\pm}(\beta)\,\mathbf{q}$ | — | $h_k^{\pm}(\beta)$ | — |
| $\tilde{\boldsymbol{w}}_k\cdot\mathbf{q}=0$ | 0 | — | — |

Table 2. Eigenvectors and eigenvalues of $\mathbf{K}$ ($\beta\neq 0$).

In particular, in order to investigate the role played by the macroscopic coupling parameter $\beta$ within the framework of the constitutive theory proposed, we can remark that the longitudinal coupled eigenvectors,

$$\left\{ \begin{array}{c} g_1^{\pm}(\beta)\,\mathbf{q} \\ g_2^{\pm}(\beta)\,\mathbf{q} \end{array} \right\}, \tag{148}$$

and their corresponding eigenvalues,

$$h_k^{\pm}(\beta) := \frac{1}{2}\left(\varrho^\star_{\rm S} c^2_{\rm LS} + \varrho^\star_{\rm F0} c^2_{\rm LF} \pm \sqrt{\left(2\,\beta\varrho^\star_{\rm F0}\right)^2 + \left(\varrho^\star_{\rm S} c^2_{\rm LS} - \varrho^\star_{\rm F0} c^2_{\rm LF}\right)^2}\right), \tag{149}$$

have to satisfy identically, for any admissible value of $\beta$ (147), the following set of algebraic equations:

$$\left\{ \begin{array}{l} \left(\varrho^\star_{\rm S} c^2_{\rm LS} - h_k^{\pm}(\beta)\right) g_1^{\pm}(\beta) - \beta\varrho^\star_{\rm F0}\, g_2^{\pm}(\beta) = 0 \\ -\beta\varrho^\star_{\rm F0}\, g_1^{\pm}(\beta) + \left(\varrho^\star_{\rm F0} c^2_{\rm LF} - h_k^{\pm}(\beta)\right) g_2^{\pm}(\beta) = 0\,, \end{array} \right. \tag{150}$$

which yields, in the general case of $\beta \neq 0$,

$$\frac{g_2^{\pm}(\beta)}{g_1^{\pm}(\beta)} = \frac{1}{2\beta\varrho^\star_{\rm F0}}\left(\varrho^\star_{\rm S} c^2_{\rm LS} - \varrho^\star_{\rm F0} c^2_{\rm LF} \mp \sqrt{\left(2\,\beta\varrho^\star_{\rm F0}\right)^2 + \left(\varrho^\star_{\rm S} c^2_{\rm LS} - \varrho^\star_{\rm F0} c^2_{\rm LF}\right)^2}\right). \tag{151}$$

† In fact, it is worth recalling that the strain-energy density per unit reference volume of the mixture (75) depends on the macroscopic kinematics of the fluid constituent, by assumption, only through the trace of its velocity gradient. Consistently, no shear wave can be sustained in the fluid.

| | $\tilde{\boldsymbol{u}}_k = \mathbf{0}$ | $\tilde{\boldsymbol{u}}_k = \mathbf{q}$ | $\tilde{\boldsymbol{u}}_k \cdot \mathbf{q} = 0$ |
|---|---|---|---|
| $\tilde{\boldsymbol{w}}_k = \mathbf{0}$ | — | $\varrho^\star_{\rm S} c^2_{\rm LS}$ | $\varrho^\star_{\rm S} c^2_{\rm TS}$ |
| $\tilde{\boldsymbol{w}}_k = \mathbf{q}$ | $\varrho^\star_{\rm F0} c^2_{\rm LF}$ | — | — |
| $\tilde{\boldsymbol{w}}_k \cdot \mathbf{q} = 0$ | 0 | — | — |

Table 3. Eigenvectors and eigenvalues of $\mathbf{K}$ ( $\beta = 0$ ).

| | $\tilde{\boldsymbol{u}}_k = \mathbf{q}$ | $\tilde{\boldsymbol{u}}_k = -g_k\,\mathbf{q}$ |
|---|---|---|
| $\tilde{\boldsymbol{w}}_k = g_k\,\mathbf{q}$ | $\varrho^\star_{\rm S} c^2_{\rm LS} + \varrho^\star_{\rm F0} c^2_{\rm LF}$ | — |
| $\tilde{\boldsymbol{w}}_k = \mathbf{q}$ | — | 0 |

Table 4. Longitudinal eigenvectors and eigenvalues of $\mathbf{K}$ ( $\beta = \beta_m$ ).

It is worth taking notice of the fact that, in the case of $\beta = 0$ (table 3),

$$\left\{ \begin{array}{l} h_k^{+}(0) = \varrho^\star_{\rm S} c^2_{\rm LS}\,, \quad \left\{ \begin{array}{c} g_1^{+}(0)\,\mathbf{q} \\ g_2^{+}(0)\,\mathbf{q} \end{array} \right\} \propto \left\{ \begin{array}{c} \mathbf{q} \\ \mathbf{0} \end{array} \right\}, \\ \\ h_k^{-}(0) = \varrho^\star_{\rm F0} c^2_{\rm LF}\,, \quad \left\{ \begin{array}{c} g_1^{-}(0)\,\mathbf{q} \\ g_2^{-}(0)\,\mathbf{q} \end{array} \right\} \propto \left\{ \begin{array}{c} \mathbf{0} \\ \mathbf{q} \end{array} \right\}, \end{array} \right. \tag{152}$$

whereas the assumption $\beta = \beta_m = (sgn\,\beta)\beta_{max}$ results in (table 4)

$$\left\{\begin{array}{ll} h_k^{+}(\beta_m) = \varrho_{\rm S}^{\star} c_{\rm LS}^2 + \varrho_{\rm F0}^{\star} c_{\rm LF}^2\,, & \left\{\begin{array}{c} g_1^{+}(\beta_m)\,\mathbf{q} \\ g_2^{+}(\beta_m)\,\mathbf{q} \end{array}\right\} \propto \left\{\begin{array}{r} \mathbf{q} \\ g_k\,\mathbf{q} \end{array}\right\}, \\ & \\ h_k^{-}(\beta_m) = 0\,, & \left\{\begin{array}{c} g_1^{-}(\beta_m)\,\mathbf{q} \\ g_2^{-}(\beta_m)\,\mathbf{q} \end{array}\right\} \propto \left\{\begin{array}{r} -g_k\,\mathbf{q} \\ \mathbf{q} \end{array}\right\}, \end{array}\right. \tag{153}$$

with

$$g_k := (sgn\,\beta_m)\sqrt{\frac{\varrho_{\rm S}^{\star} c_{\rm LS}^2}{\varrho_{\rm F0}^{\star} c_{\rm LF}^2}} = \frac{sgn\,\beta_m}{\varrho_{\rm F0}^{\star}}\sqrt{\frac{\lambda_{\rm S} + 2\mu_{\rm S} + \alpha}{b}}\,; \quad sgn\,\beta_m = \frac{\beta_m}{|\,\beta_m\,|}\,. \tag{154}$$

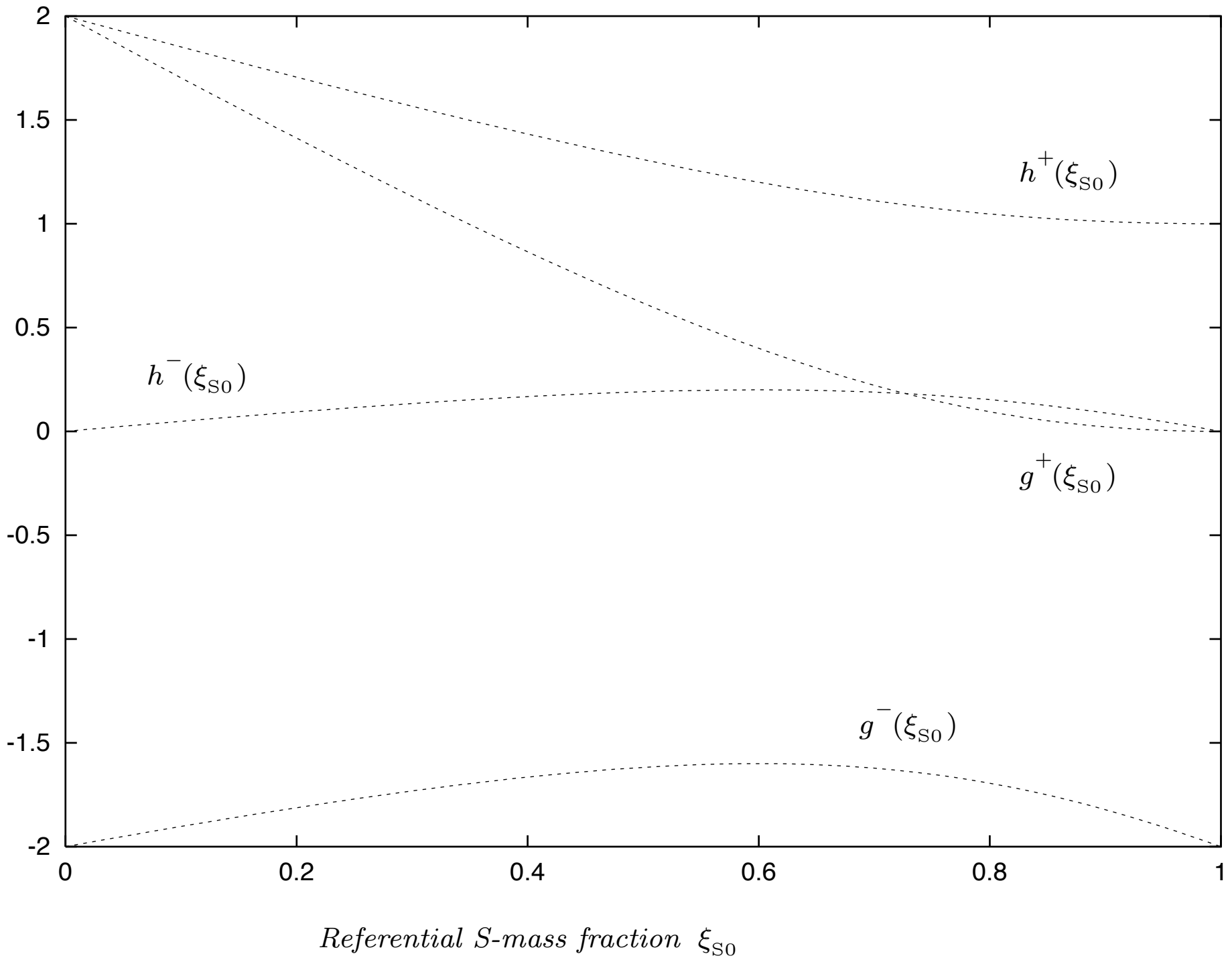

Figure 2. Eigenvectors and eigenvalues of $\mathbf{M}$ ( $h^{\pm}(\xi_{\rm S0})$ and $g^{\pm}(\xi_{\rm S0})$) .

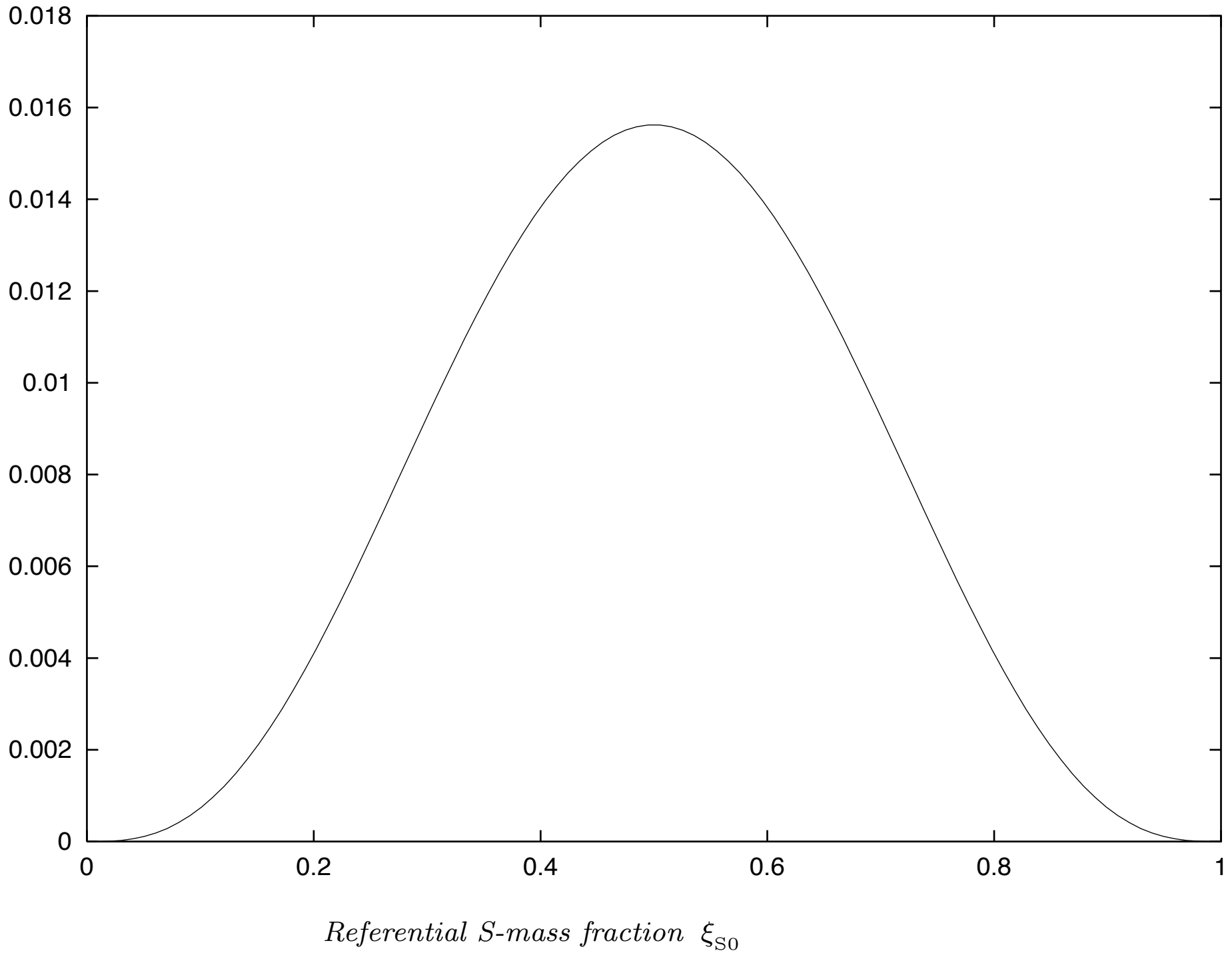

Figure 3. $\dfrac{\det \mathbf{M}}{{\varrho_0^\star}^6} = (\xi_{F0}\, \xi_{S0})^3$ .

We can now investigate the overall dynamical properties of the mixture, emerging from the analysis of the algebraic compact form (139) of the set of differential equations (132)-(133),

$$\begin{aligned} \left(\varrho_S^\star c_{TS}^2 - \varrho_0^\star c^2\right) \tilde{\boldsymbol{u}}_c + \varrho_S^\star \left(c_{LS}^2 - c_{TS}^2\right) (\mathbf{q}\cdot\tilde{\boldsymbol{u}}_c)\, \mathbf{q} \,+ & \\ -\beta \varrho_{F0}^\star (\mathbf{q}\cdot\tilde{\boldsymbol{w}}_c)\, \mathbf{q} - \varrho_{F0}^\star c^2\, \tilde{\boldsymbol{w}}_c &= 0 \qquad (155) \\ -\varrho_{F0}^\star c^2\, \tilde{\boldsymbol{u}}_c - \varrho_{F0}^\star c^2\, \tilde{\boldsymbol{w}}_c - \beta \varrho_{F0}^\star (\mathbf{q}\cdot\tilde{\boldsymbol{u}}_c)\, \mathbf{q} + \varrho_{F0}^\star c_{LF}^2 (\mathbf{q}\cdot\tilde{\boldsymbol{w}}_c)\, \mathbf{q} &= 0\,, \qquad (156) \end{aligned}$$

whose solutions physically characterize all possible kinds of plane elastic waves that can be sustained in the medium, according to the proposed constitutive (macroscopic) theory.

**Transverse waves.** As the strain-energy density per unit reference volume of the mixture is assumed to depend on the macroscopic kinematics of the fluid constituent only through the trace of its velocity gradient, the eigenvectors

$$\left\{ \begin{array}{c} \tilde{\boldsymbol{u}}_c \\ \tilde{\boldsymbol{w}}_c \end{array} \right\} \propto \left\{ \begin{array}{c} \mathbf{0} \\ \tilde{\boldsymbol{w}}_c \end{array} \right\}, \qquad \text{with} \qquad \tilde{\boldsymbol{w}}_c \cdot \mathbf{q} = 0\,, \tag{157}$$

are naturally associated with null eigenvalues. Moreover, it can be straightforwardly deduced from equations (155)-(156) that elastic *transverse* waves are associated with the eigenvectors that satisfy the requirements:

$$\tilde{\boldsymbol{u}}_c \cdot \mathbf{q} = \tilde{\boldsymbol{w}}_c \cdot \mathbf{q} = 0\,, \tag{158}$$

$$\tilde{\boldsymbol{u}}_c + \tilde{\boldsymbol{w}}_c = \mathbf{0}\,. \tag{159}$$

By virtue of constitutive prescriptions assumed for the coupling coefficient $\mathbf{B}$ (124), the resulting phase velocity of transverse waves propagating in the mixture, $c = c_{\text{T}}$, is exactly equal to the characteristic speed of transverse waves that would propagate in the solid constituent, $c_{\text{T}} = c_{\text{TS}}$.

**Longitudinal waves.** Longitudinal coupled eigenvectors,

$$\left\{ \begin{array}{c} \tilde{\boldsymbol{u}}_c \\ \tilde{\boldsymbol{w}}_c \end{array} \right\} \propto \left\{ \begin{array}{c} \lambda_1^{\pm}(\beta)\,\mathbf{q} \\ \lambda_2^{\pm}(\beta)\,\mathbf{q} \end{array} \right\}, \tag{160}$$

are naturally associated with the longitudinal waves that can be sustained in the mixture, whose phase velocities

$$\left(c_{\text{L}}^2\right)^{\pm} = \frac{1}{2\varrho_{\text{S}}^{\star}} \left\{ h(\beta) \pm \sqrt{h^2(\beta) + 4\varrho_{\text{S}}^{\star}\varrho_{\text{F0}}^{\star}\left(\beta^2 - \beta_{max}^2\right)} \right\}, \tag{161}$$

with (cf. Biot [3], Wilmanski [38], Edelman and Wilmanski [14])

$$h(\beta) = \varrho_{\text{S}}^{\star} c_{\text{LS}}^2 + \left(\varrho_{\text{F0}}^{\star} + \varrho_{\text{S}}^{\star}\right) c_{\text{LF}}^2 + 2\beta\varrho_{\text{F0}}^{\star}\,, \tag{162}$$

are given by the solutions of the characteristic equation

$$\xi_{\text{F0}}\left(c_{\text{L}}^2 + \beta\right) - \left(c_{\text{L}}^2 - c_{\text{LF}}^2\right)\left(c_{\text{L}}^2 - \xi_{\text{S0}} c_{\text{LS}}^2\right) = 0\,. \tag{163}$$

Accordingly, the coefficients $\lambda_1^{\pm}(\beta)$ and $\lambda_2^{\pm}(\beta)$ in (160) have to satisfy identically, for any admissible value of $\beta$ (147), the set of algebraic equations:

$$\left\{ \begin{array}{l} \left\{\varrho_{\text{S}}^{\star} c_{\text{LS}}^2 - \left(\varrho_{\text{S}}^{\star} + \varrho_{\text{F0}}^{\star}\right)\left(c_{\text{L}}^2\right)^{\pm}\right\} \lambda_1^{\pm}(\beta) - \left\{\beta\varrho_{\text{F0}}^{\star} + \varrho_{\text{F0}}^{\star}\left(c_{\text{L}}^2\right)^{\pm}\right\} \lambda_2^{\pm}(\beta) = 0 \\ -\left\{\beta\varrho_{\text{F0}}^{\star} + \varrho_{\text{F0}}^{\star}\left(c_{\text{L}}^2\right)^{\pm}\right\} \lambda_1^{\pm}(\beta) + \left\{\varrho_{\text{F0}}^{\star} c_{\text{LF}}^2 - \varrho_{\text{F0}}^{\star}\left(c_{\text{L}}^2\right)^{\pm}\right\} \lambda_2^{\pm}(\beta) = 0\,. \end{array} \right. \tag{164}$$

## Constrained solid-fluid mixture

Although it proves itself capable to catch the bare essentials of the mechanical behavior exhibited by a poroelastic solid infused with a Stokesian fluid (cf. Cowin [8]), the theoretical model presented so far is purely *macroscopic.*

To take into account the most relevant microstructural properties of the mixture, we may better enrich such a model by introducing two independent scalar fields of volume fraction (see sections 4 and 5.2), constrained by the kinematical requirement (43). As kinematical constraints are naturally associated with reactive actions, a *saturation pressure* $p(\mathbf{X},t)$ needs to arise in the mixture so as to maintain each constituent in contact with the other one. Regarding the saturation pressure as a Lagrangian multiplier in the expression of the overall strain-energy density per unit reference volume (100), and recalling that, by assumption,

$$\operatorname{Grad} \varrho^\star_{\mathrm{F0}} = \mathbf{0} \tag{165}$$

and (compare with expressions (128)-(130)),

$$\begin{aligned}
\mathbf{T}_1(t) &= \left\{\beta \varrho^\star_{\mathrm{F1}}(t) + \lambda_{\mathrm{S}} \operatorname{Div} \boldsymbol{u}_{\mathrm{S}}(t) + p_1(t)\,\beta_{\mathrm{S}}\right\}\mathbf{I} + \\
&\quad + 2\mu_{\mathrm{S}} \operatorname{sym}(\operatorname{Grad} \boldsymbol{u}_{\mathrm{S}}(t)) + \alpha \operatorname{Grad} \boldsymbol{u}_{\mathrm{S}}(t) &(166)\\
\boldsymbol{\tau}_1(t) &= \operatorname{Grad}\left\{\xi_{\mathrm{F1}}(t)\,\widehat{\mathfrak{W}}_o + \xi_{\mathrm{F0}}\,(\mathbf{A} + p_0\beta_{\mathrm{S}}\mathbf{I}) \cdot \operatorname{Grad} \boldsymbol{u}_{\mathrm{S}}(t)\right\} + \\
&\quad + \operatorname{Grad}\left\{\xi_{\mathrm{F0}}\left(a + \beta_{\mathrm{F}}\frac{p_0}{\hat{\varrho}_{\mathrm{F0}}}\right)\varrho^\star_{\mathrm{F1}}(t)\right\} &(167)\\
\boldsymbol{\mathcal{T}}_1(t) &= \left\{\xi_{\mathrm{F1}}(t)\,\widehat{\mathfrak{W}}_o + \xi_{\mathrm{F0}}\left(a + \beta_{\mathrm{F}}\frac{p_0}{\hat{\varrho}_{\mathrm{F0}}}\right)\varrho^\star_{\mathrm{F1}}(t)\right\}\mathbf{I} + \\
&\quad + \left\{\xi_{\mathrm{F0}}\,(\mathbf{A} + p_0\beta_{\mathrm{S}}\mathbf{I}) \cdot \operatorname{Grad} \boldsymbol{u}_{\mathrm{S}}(t)\right\}\mathbf{I} + \\
&\quad - \varrho^\star_{\mathrm{F0}}\left\{b\varrho^\star_{\mathrm{F1}}(t) + \beta \operatorname{Div} \boldsymbol{u}_{\mathrm{S}}(t)\right\}\mathbf{I}, &(168)
\end{aligned}$$

it is possible to formulate the problem in terms of the field quadruplet $\{\boldsymbol{u}_{\mathrm{S}}, \boldsymbol{w}_{\mathrm{F}}, \varrho^\star_{\mathrm{F1}}, p\}$,

$$\begin{aligned}
\frac{\partial \varrho^\star_{\mathrm{F1}}}{\partial t} + \varrho^\star_{\mathrm{F0}} \operatorname{Div} \boldsymbol{w}_{\mathrm{F}} &= 0 &(169)\\
\left(\varrho^\star_{\mathrm{S}} + \varrho^\star_{\mathrm{F0}}\right)\frac{\partial^2 \boldsymbol{u}_{\mathrm{S}}}{\partial t^2} + \varrho^\star_{\mathrm{F0}}\frac{\partial \boldsymbol{w}_{\mathrm{F}}}{\partial t} - (\lambda_{\mathrm{S}} + \mu_{\mathrm{S}}) \operatorname{Grad}(\operatorname{Div} \boldsymbol{u}_{\mathrm{S}}) + \\
- (\mu_{\mathrm{S}} + \alpha) \operatorname{Div}(\operatorname{Grad} \boldsymbol{u}_{\mathrm{S}}) - \beta \operatorname{Grad} \varrho^\star_{\mathrm{F1}} - \beta_{\mathrm{S}} \operatorname{Grad} p_1 &= \mathbf{0} &(170)\\
\varrho^\star_{\mathrm{F0}}\frac{\partial^2 \boldsymbol{u}_{\mathrm{S}}}{\partial t^2} + \varrho^\star_{\mathrm{F0}}\frac{\partial \boldsymbol{w}_{\mathrm{F}}}{\partial t} + \varrho^\star_{\mathrm{F0}}\,\beta \operatorname{Grad}(\operatorname{Div} \boldsymbol{u}_{\mathrm{S}}) + \varrho^\star_{\mathrm{F0}}\, b \operatorname{Grad} \varrho^\star_{\mathrm{F1}} + \\
+ \nu_{\mathrm{F0}}\beta_{\mathrm{F}} \operatorname{Grad} p_1 &= \mathbf{0} &(171)\\
\hat{\varrho}_{\mathrm{F0}}\,\beta_{\mathrm{S}} \operatorname{Div} \boldsymbol{u}_{\mathrm{S}}(t) + \beta_{\mathrm{F}}\,\varrho^\star_{\mathrm{F1}}(t) &= 0, &(172)
\end{aligned}$$

whereas the (unperturbed) reference state is characterized by a further set of linear equations, deducible from $(121)_1$,

$$\left\{\begin{aligned}
&\operatorname{Div} \mathbf{T}_0 = \operatorname{Div}(\alpha + p_0\beta_{\mathrm{S}})\,\mathbf{I} = \mathbf{0}\\
&\operatorname{Div} \boldsymbol{\mathcal{T}}_0 - \boldsymbol{\tau}_0 = -\varrho^\star_{\mathrm{F0}} \operatorname{Grad}\left(a + \beta_{\mathrm{F}}\frac{p_0}{\hat{\varrho}_{\mathrm{F0}}}\right) = \mathbf{0}.
\end{aligned}\right. \tag{173}$$

Looking for steady-state solutions in the form:

$$\boldsymbol{u}_{\mathrm{S}}(\mathbf{X},t) = \tilde{\boldsymbol{u}}\, e^{ik(\mathbf{X}\cdot\,\mathbf{q}-ct)} \tag{174}$$

$$\boldsymbol{w}_{\mathrm{F}}(\mathbf{X},t) = -\,ikc\,\tilde{\boldsymbol{w}}\, e^{ik(\mathbf{X}\cdot\,\mathbf{q}-ct)} \tag{175}$$

$$\varrho^{\star}_{\mathrm{F1}}(\mathbf{X},t) = \tilde{\varrho}\, e^{ik(\mathbf{X}\cdot\,\mathbf{q}-ct)} \tag{176}$$

$$p_{1}(\mathbf{X},t) = \tilde{p}\, e^{ik(\mathbf{X}\cdot\,\mathbf{q}-ct)}\,, \tag{177}$$

we notice that, by virtue of the local fluid-mass conservation law (169) and the kinematical constraint (172),

$$\tilde{\varrho} = -\,ik\varrho^{\star}_{\mathrm{F0}}\,(\tilde{\boldsymbol{w}}\cdot\mathbf{q}) \tag{178}$$

$$\tilde{\varrho} = -\,ik\,\frac{\beta_{\mathrm{S}}}{\beta_{\mathrm{F}}}\,\varrho_{\mathrm{F0}}\,(\tilde{\boldsymbol{u}}\cdot\mathbf{q})\,, \tag{179}$$

the reduced set of equations (170)-(172) can be uncoupled from the fluid-mass conservation law (169). Moreover, as the gradient of the pressure field $p_{1}(t)$ is indeed parallel to the direction of wave propagation (defined by the unit vector $\mathbf{q}$), the set of equations that describes the longitudinal dynamics in terms of the scalar unknowns $\{\tilde{\varrho},\tilde{p},\tilde{u}_{\mathrm{L}},\tilde{w}_{\mathrm{L}}\}$ can be uncoupled from the one that describes the transversal dynamics in terms of $\{\tilde{\boldsymbol{u}}_{\mathrm{T}},\,\tilde{\boldsymbol{w}}_{\mathrm{T}}\}$, where

$$\tilde{\boldsymbol{u}} := \tilde{u}_{\mathrm{L}}\mathbf{q}+\tilde{\boldsymbol{u}}_{\mathrm{T}}\,,\qquad \tilde{\boldsymbol{u}}_{\mathrm{T}}\cdot\mathbf{q}=0 \tag{180}$$

$$\tilde{\boldsymbol{w}} := \tilde{w}_{\mathrm{L}}\mathbf{q}+\tilde{\boldsymbol{w}}_{\mathrm{T}}\,,\qquad \tilde{\boldsymbol{w}}_{\mathrm{T}}\cdot\mathbf{q}=0\,. \tag{181}$$

Accordingly, it is worth emphasizing that the transverse dynamics is *unaffected* by the saturation constraint (172).

**Longitudinal waves** Looking for longitudinal steady-state solutions of equations (169)-(172) and combining, respectively, the fluid-mass conservation law (169) with the kinematical constraint (172), and Cauchy's law of motion (170) with the analogous equation (171), we can at first focus our attention on a reduced set of scalar (algebraic) equations in the field doublet $\{\tilde{u}_{\mathrm{L}},\tilde{w}_{\mathrm{L}}\}$,

$$\begin{cases} \mathrm{D}_{11}\tilde{u}_{\mathrm{L}}+\mathrm{D}_{12}\tilde{w}_{\mathrm{L}}=0\\ \mathrm{D}_{21}\tilde{u}_{\mathrm{L}}+\mathrm{D}_{22}\tilde{w}_{\mathrm{L}}=0\,, \end{cases} \tag{182}$$

with

$$\begin{cases} \mathrm{D}_{11} := \beta_{\mathrm{S}}\\ \mathrm{D}_{12} := -\nu_{\mathrm{F0}}\beta_{\mathrm{F}}\\ \mathrm{D}_{21} := \nu_{\mathrm{F0}}\beta_{\mathrm{F}}\varrho^{\star}_{\mathrm{S}}c^{2}_{\mathrm{LS}}-\beta_{\mathrm{S}}\beta\varrho^{\star}_{\mathrm{F0}}-\left(\nu_{\mathrm{F0}}\beta_{\mathrm{F}}\left(\varrho^{\star}_{\mathrm{S}}+\varrho^{\star}_{\mathrm{F0}}\right)+\beta_{\mathrm{S}}\varrho^{\star}_{\mathrm{F0}}\right)c^{2}\\ \mathrm{D}_{22} := \beta_{\mathrm{S}}\varrho^{\star}_{\mathrm{F0}}c^{2}_{\mathrm{LF}}-\nu_{\mathrm{F0}}\beta_{\mathrm{F}}\beta\varrho^{\star}_{\mathrm{F0}}-\varrho^{\star}_{\mathrm{F0}}\left(\beta_{\mathrm{S}}+\nu_{\mathrm{F0}}\beta_{\mathrm{F}}\right)c^{2}\,. \end{cases} \tag{183}$$

As the coefficients $D_{11}$ and $D_{12}$ do not depend on the longitudinal characteristic squared speed $c^2 = c^2_{L}$, the equation

$$D_{11}D_{22} - D_{12}D_{21} = 0 \tag{184}$$

is linear in $c^2$, and yields the unique solution

$$c^2 = c^2_{L} = \frac{\varrho^\star_{S} c^2_{LS} \left(\nu_{F0}\beta_{F}\right)^2 - 2\beta_{S}\left(\nu_{F0}\beta_{F}\right)\left(\beta\varrho^\star_{F0}\right) + \varrho^\star_{F0} c^2_{LF}\left(\beta_{S}\right)^2}{\varrho^\star_{F0}\left(\nu_{F0}\beta_{F} + \beta_{S}\right)^2 + \varrho^\star_{S}\left(\nu_{F0}\beta_{F}\right)^2} . \tag{185}$$

Henceforth, a dependence of the characteristic longitudinal speed on the macroscopic coupling parameter $\beta$ (124) can be taken into account in the case of fluid-saturated poroelastic solids (185) as well as in the case of unconstrained solid-fluid mixtures (161). In particular, recalling the requirement $|\beta| \leq \beta_{max}$, we may finally point out, by virtue of the linear dependence of $c^2_{L}$ on $\beta$, that the square of the characteristic speed of the (unique) longitudinal plane wave (185) has to range from

$$\left(c^2_{L}\right)_{min} = \frac{\left\{ \nu_{F0}\beta_{F}\sqrt{\varrho^\star_{S} c^2_{LS}} - \beta_{S}\sqrt{\varrho^\star_{F0} c^2_{LF}} \right\}^2}{\varrho^\star_{F0}\left(\nu_{F0}\beta_{F} + \beta_{S}\right)^2 + \varrho^\star_{S}\left(\nu_{F0}\beta_{F}\right)^2} \qquad (\beta = \beta_{max}) , \tag{186}$$

to

$$\left(c^2_{L}\right)_{max} = \frac{\left\{ \nu_{F0}\beta_{F}\sqrt{\varrho^\star_{S} c^2_{LS}} + \beta_{S}\sqrt{\varrho^\star_{F0} c^2_{LF}} \right\}^2}{\varrho^\star_{F0}\left(\nu_{F0}\beta_{F} + \beta_{S}\right)^2 + \varrho^\star_{S}\left(\nu_{F0}\beta_{F}\right)^2} \qquad (\beta = -\beta_{max}) , \tag{187}$$

whenever the saturation constraint is kinematically satisfied.

In conclusion, we remark that the presence of such a constraint (43), that obviously reduces the degree of freedom of the longitudinal dynamics of the mixture, furthermore allows the microstructural constitutive parameters $\beta_{S}$ and $\beta_{F}$ to contribute to the resulting characteristic speed $c_{L}$ (refer to definitions $(51)_1$ and $(51)_2$, formerly introduced in section 4).

Accordingly, a relevant dependence on the constitutive information associated with the definition of microscopic mass-density fields (45)-(46) may be brought forth at the macroscopic level.

## Acknowledgments

This work has been developed within the framework of the TMR European Network on "Phase Transitions in Crystalline Solids". Fruitful discussions with Prof. Daniel Lhuillier are gratefully acknowledged.

Sara Quiligotti
Laboratoire de Modélisation en Mécanique
Université Pierre et Marie Curie
4 place Jussieu, 75252 Paris, FRANCE
(quiligotti@lmm.jussieu.fr)

Gérard A. Maugin
Laboratoire de Modélisation en Mécanique
Université Pierre et Marie Curie
4 place Jussieu, 75252 Paris, FRANCE
(gam@ccr.jussieu.fr)

Francesco dell'Isola
Dipartimento di Ingegneria Strutturale e Geotecnica
Università degli Studi di Roma "La Sapienza"
Via Eudossiana 18, 00184 Roma, ITALY
(francesco.dellisola@uniroma1.it)